\documentclass[a4paper,11pt,epsfig]{article}
\usepackage{epsfig}
\newcommand{\be}{\begin{equation}}
\newcommand{\ee}{\end{equation}}
\newcommand{\ba}{\begin{eqnarray}}
\newcommand{\ea}{\end{eqnarray}}
\newcommand{\nn}{\nonumber}

\newcommand{\namu}{\nabla_\mu}
\newcommand{\nanu}{\nabla_\nu}

\newcommand{\av}[1]{{\mathrm{av}}_{#1}}

\newcommand{\Aop}{\mathcal{A}}
\newcommand{\beff}{\beta_{\mathrm{eff}}}
\newcommand{\Cav}{\mathcal{C}}
\newcommand{\Lap}{\triangle}
\newcommand{\einhalb}{\frac{1}{2}}

\newcommand{\Seff}{S_\mathrm{eff}}
\newcommand{\tr}{\mathrm{tr}\,}

\newcommand{\TSz}{T_{\Psi (z)}\mathcal{S}^{\mathcal{N}}}

\newcommand{\TSx}{T_{\Phi (x)}\mathcal{S}^{\mathcal{N}}} 

\newcommand{\RNP}{ \mathbf{R}^{\mathcal{N}+1} }  
\newcommand{\PsI}{ \underline{\Psi }}
\newcommand{\PhI}{ \underline{\Phi }}
\newcommand{\phI}{ \underline{\phi }}
\newcommand{\xI}{ \underline{\xi }}

\newcommand{\leftNabla}{\stackrel{\leftarrow}{\nabla}}

\title{Analytic Calculation of the 1-loop effective action
 for the O($\mathcal{N}+1$)-symmetric 2-dimensional nonlinear
$\sigma$-model\footnote{Work supported by Deutsche Forschungsgemeinschaft}
}

\author{M.\,Bartels, G.\,Mack \\
        II.\,Institut f\"ur Theoretische Physik der Universit\"at
Hamburg,\\
        D-22761 Hamburg, Luruper Chaussee 149, Germany\\
         \\
         G.\,Palma\\
         Departamento Fisica, Universidad de Santiago de Chile,\\
         Casilla 307, Correo 2, Santiago, Chile
         }

\begin{document}

\maketitle

\begin{abstract}

Starting from the 2-dimensional nonlinear
 $\sigma$-model living on a lattice $\Lambda$ of lattice spacing $a$
with action
$S[\phi]=-\einhalb\beta\int_z\,\phi\Lap\phi,\quad \phi(z)\in
S^\mathcal{N}$ we
compute the Wilson effective action $S_\mathrm{eff}[\Phi]$ on a lattice of
lattice spacing $\tilde a$
in a 1-loop approximation for a choice of blockspin
$\Phi(x)=  C\phi(x)\equiv \Cav\phi(x)/{|\Cav\phi(x)|}$,
where $\Cav$ is averaging of $\phi(z)$ over a block $x$. We use a
$\delta$-function constraint to enforce $\Phi = C \phi $.
We consider also
a Gaussian in place of the $\delta$-function
in order to improve locality properties of $S_\mathrm{eff}$
 as proposed by Hasenfratz
 and Niedermayer.

The result for $S_\mathrm{eff}$ is composed of
the classical perfect action with a renormalized coupling
 constant $\beff$, an augmented contribution from a Jacobian,
 and further correction terms.
The jacobian term depends on $\Psi(z)\cdot \Phi(x)$
where $\Psi $ is the interpolation of $\Phi $ with minimal action.
 The further correction terms include $\Psi\cdot \Phi$-dependent
 fluctuations of $\beff$ and a genuine 1-loop correction which
depends on the matrix $\Psi \nabla_\mu \Psi^T(z)$ at two different sites
$z=z_1,z_2$.  We find an analytic approximation for $\Psi $. Using it
one can express the  classical perfect action and $S_\mathrm{eff}$ as a
function of the block spin $\Phi$.
Our result extends Polyakovs calculation which had furnished
those contributions to the effective action which are of order
$\ln \tilde a /a$.

\end{abstract}

\section{Introduction. Perfect actions}

\label{sec:perfact}

Perfect actions are actions for a lattice field theory which reproduce the
expectation values of a continuum theory or of a theory with a much higher
UV
cutoff for a restricted class of ``low energy''observables.
Effective lattice actions in the sense of Wilson are perfect actions in
this
sense.
Different approximations to the Wilson effective action have been given
names
such as classical perfect actions, 1-loop perfect actions etc. 
\cite{HN94,H97}.

In this paper we compute the effective lattice action for the
2-dimensional
$O(\mathcal{N}+1)$-symmetric nonlinear $\sigma$-model in a 1-loop
approximation. 
The result is given in eqs.(\ref{result})ff below. A summary was presented in 
\cite{BMPPisa}.

The model lives on a quadratic lattice $\Lambda$ of lattice spacing $a$
with
points typically denoted  $z,w,\dots .$
Let $\hat{\mu}$ be the lattice vector of length $a$ in $\mu$-direction
($\mu=1,2$).
We use lattice notations as follows (and similarly for the block lattice
below).
\ba
   \int_z (\dots )&=&a^2\sum_{z\in\Lambda} (\dots );
\\ \qquad \namu f(z)&=&\frac{1}{a}[f(z+\hat{\mu})-f(z)]
= - \nabla_{-\mu}f(z+\hat \mu);\\
\qquad -\Lap&=&\sum_{\mu=1,2}\namu\nabla_{-\mu} \ .
  \label{latnot3}
\ea
In the continuum limit $a\to 0$, $\int_z\to\int d^2z$.
The field $\phi(z)\in S^\mathcal{N}$ is a ($\mathcal{N}+1$)-dimensional
unit
vector, and $d\phi$ is the normalized uniform measure on the sphere.
The action of the model is
\begin{equation}
  \label{act}
S[\phi]=
\frac{\beta}{2}\int_z[\namu\phi(z)]^2=-\frac{\beta}{2}\int_z\phi\Lap\phi .
\end{equation}
A block lattice $\tilde{\Lambda}$ of lattice spacing $\tilde{a}=s\cdot a$ is
superimposed ($s$ positive integer).
Its points are typically denoted  $x,y,\dots .$
They are identified with squares of sidelength $\tilde{a}$ in $\Lambda$.
There are $s^2$ points $z\in x$.

We define a blockspin $\Phi(x)$ which lives on the block lattice as a
function
$\Phi(x)=C\phi(x)$ of the fundamental field.
$\Phi(x)$ is also a ($\mathcal{N}$+1)-unit vector; therefore the operator
$C$
is necessarily nonlinear.
We choose
\begin{equation}
  \label{Cop}
  \Phi(x)=C\phi(x)\equiv\frac{\Cav\phi(x)}{|\Cav\phi(x)|} \ , 
\end{equation}
where $\Cav$ is a linear operator which averages over blocks.
We take
\begin{equation}
  \label{Cav}
  \Cav\phi(x)=\av{z\in x}\phi(z)\equiv \tilde{a}^{-2}\int_{z\in x} \phi(z).
\end{equation}
The Wilson effective action is defined by
\ba
  \label{weact}
e^{-S_\mathrm{eff}[\Phi]}&=&
        \int\,\mathcal{D}\phi\prod_x\delta(C\phi(x),\Phi(x))e^{-S[\phi]}\ ;\\
\mathcal{D}\phi &=&\prod_z d\phi(z) ,
\ea
where $d\phi $ is the uniform measure on the sphere $S^\mathcal{N}$,
and $\delta$ is the $\mathcal{N}$-dimensional $\delta$-function on the
sphere,
viz.
\begin{displaymath}
  \int\,d\phi_1 f(\phi_1)\delta(\phi_1,\phi_2)=f(\phi_2)
\end{displaymath}
for test functions on the sphere.

We use a $\delta$-function constraint
 because computation
of expectation values of observables which depend on $\phi $ only through
the
blockspin $\Phi $ must then be identical whether computed with $S$ or
$\Seff$.
 This prepares best for stringent tests of the accuracy of the result.

Hasenfratz and Niedermayer \cite{HN94} 
showed numerically that much better locality 
properties of effective actions are obtained when a Gaussian is used 
in the definition of the effective action in place 
of a sharp $\delta$-function. 

 This motivates us to examine also a 
larger class of block spin transformations which depends on a parameter
 $\kappa $ and 
which use a Gaussian in place of a $\delta$-function.
The $\delta$-function constraint is obtained in the limit
 $\kappa  \mapsto \infty $.

The calculation proceeds in the same way as in the $\kappa = \infty$ case,
and the result is also the same, except
\begin{itemize}
\item The interpolation kernels $\Aop $ and high frequency propagators $\Gamma $
with finite $\kappa $ must be used throughout.
\item
The background field $\Psi $ is determined by eq.(\ref{extrKappa}) of section 
\ref{sec:gaussian}, but the analytic expression for $\Psi$ remains valid 
when the finite-$\kappa$ expressions for $\Aop$ and $\Gamma$ are used.  

\item the classical perfect action differs from $S(\Psi ) $ by an extra term
\hfill \\
$\frac 12 \beta \kappa \int_x |\Cav \Psi^\perp (x)|^2, $ where 
$\Psi^\perp (z) = \Psi (z) - \Phi (x) \left( \Psi(z) \cdot \Phi (x)\right) $
 for $z\in x$.
\item the 
  jacobian $J_0$ receives extra contributions of order $\kappa^{-1}$.
\end{itemize}

Details are presented in section \ref{sec:gaussian}. 

The analytic formula for the background field $\Psi $
elucidates the better locality properties of the effective action for suitable
finite $\kappa$. It comes from the better locality properties of the
Kupiainen Gawedzki high frequency propagator $\Gamma_{KG}$.             

\subsection*{Background field and classical perfect action}

Given a blockspin configuration $\Phi$, let $\Psi=\Psi[\Phi]$ be that
field on
the fine lattice $\Lambda$ which extremizes $S(\phi ) + \frac {\tilde \kappa}{2} \sum_x |\Cav \phi^\perp (x)|^2  $ subject to the
constraints $|\Psi (z)|^2 = 1 $ and
\begin{displaymath}
  C\Psi=\Phi
\label{Psi}
\end{displaymath}
$\Psi$ is called the background field.
The  classical perfect action is
\begin{equation}
  \label{classperfact}
  S_{cl}[\Phi ]=S(\Psi[\Phi]).
\end{equation}
Here we wish to compute the 1-loop corrections.
It is convenient to regard the full effective action as a function of
$\Psi $. This is possible because
$\Phi $ is determined by $\Psi$ according to eq.(\ref{Psi}).

For large enough blocks, the background field $\Psi $ is smooth.  An
 analytical  approximation for $\Psi $ as a function of $\Phi $ is derived in
 section \ref{sec:background}.

\subsection*{Summary of results}
Because of the smoothness of $\Psi$, it sufficies to consider terms up 
to second order in $\nabla \Psi$. 
The exact 1-loop perfect action to this  order  is as follows.
\footnote{To save brackets, we adopt the notational convention that a
 derivatives acts only on the factor immediately following it.
We used vector notation, $\Psi^T$ is the row vector transpose to $\Psi$.
Note that $j_\mu(z)$ is a matrix.}
\ba
\Seff &=&  S_{cl} - \sum_x\ln J_0(\Cav \Psi (x)) 
- \frac 12 Tr \ln \Gamma_Q  \nn  \\
&& + \frac 12\int \left(
\nabla_\mu \Psi^T(z) \beff^1(z)\nabla_\mu \Psi(z) 
+
\beff^2(z) \frac  {\Phi([z])^T (-\Lap)\Psi (z)} {\cos\theta (z)}
\right) \nn  \\
& & + \Seff^{(2)}
+ \int_z \tr j_\mu(z)\nabla_\mu \Gamma_Q (z, w)|_{w=z+\hat \mu}  \
 ; \label{result} \\
j_\mu (z) &=& \Psi(z)\nabla_\mu \Psi^T(z)  - \nabla_\mu\Psi(z)\Psi^T(z+\hat{\mu}) 
+ \Psi \Psi^T \nabla_{\mu} \Psi \Psi^T(z+\hat \mu )   \ ,
\label{alphaDef}
\ea
where $[z]$ is the block containing $z$, the jacobian is
\be
J_0(\Cav \Psi (x) ) = | \Phi\cdot\Cav\Psi (x)|^\mathcal{N} \label{jacIntro}
\ee
and $S_\mathrm{eff}^{(2)} $ is a 
contribution from a renormalized 1-loop graph with 
2 vertices as follows
\ba
\Seff^{(2)} &=& 
-\frac 12 \int_z\int_w \tr \left( \nabla_\mu \Gamma_Q(z,w) \leftNabla_\nu 
j_\nu^T(w)\Gamma_Q(w,z) j_\mu(z)  \right. \nn \\
&& \qquad \qquad 
+ \nabla_\mu \Gamma_Q (z,w) j_\nu(w) \nabla_\nu \Gamma_Q (w,z) 
j_\mu(z)  \nn \\
& &  \left. \qquad \qquad + \
 \delta_{\mu \nu}\delta(z-w) j_\mu(z)\Gamma_Q(z+\hat \mu,z+ \hat \mu)j_\nu(z)
\right)  \ .
\label{ren1loop}
\ea
The $\delta_{\mu \nu}\delta(z-w) $-term subtracts the part which diverges
in the limit $a\mapsto 0$. The last term in the definition 
(\ref{alphaDef}) of $j_\mu $ is a lattice artifact and 
can be dropped inside eq.(\ref{ren1loop})
because its contribution is actually of higher order  in $\nabla \Psi$.   

$\Gamma_Q$ is an $(\mathcal{N}+1) \times (\mathcal{N}+1)$ matrix 
propagator,  
\be
\Gamma_Q = \lim_{\kappa\mapsto \infty }
(-\Lap + \kappa \hat Q^T \Cav^\dagger \Cav \hat Q )^{-1}
\ee
with 
\be
\hat Q(z)  = 1 - \Psi (z)\Psi^T (z) + 
\Phi (x) (\Psi^T (z)[1+ \cos \theta (z)] - \Phi^T (x))\ , \label{Qhat0} 
\ee
and
\be
\cos \theta (z) = \Psi (z)\cdot \Phi (x)\ , \ (x\ni z).
\ee
The coupling constant renormalizations $\beff^1$ and $\beff^2$ both have a 
residual dependence on $\Psi $ through $\hat Q$,
so they fluctuate somewhat with $\Psi$; to leading order 
the dependence is through $\cos\theta$. Note that  $\beff^1$ is a 
$(\mathcal{N}+1) \times (\mathcal{N}+1)$ matrix,
 while $\beff^2 $ is a scalar. 
\ba
\beff^1(z) &=& \Gamma_Q(z,z)\ , \\ 
\beff^2(z) &=& -\tr [1-\Psi \Psi^T(z)]\Gamma_Q(z,z)\ .
\ea  
Finally, the last  term in eq.(\ref{result}) is a  lattice artifact; 
cp. Appendix B. and below.

Because of the complicated $\Psi$-dependence of the propagators, 
the exact result of the 1-loop calculation is too complicated to be of much
practical use. Simple approximations require additional assumptions to
justify them. 

Assuming a  smooth enough block spin field $\Phi (x)$, the matrix 
$\hat Q$ is close to $1$ and we may expand in powers of $\hat Q -1$.
It will be shown in section \ref{sec:fieldDep} how to compute the corrections. 

To leading order, the terms of order $\hat Q$ -1 will be neglected except 
in the $Tr \ln \Gamma_Q $-term, using
\be
\Gamma_Q = \Gamma_{KG}{\bf 1}  + O(\hat Q -1),
\ee
where $\Gamma_{KG}$ is the Kupiainen Gawedzki high frequency propagator 
for scalar fields as defined below. Splitting
\be
[\cos \theta (z)  ]^{-1}\Phi (x) = \Psi + [\cos \theta (z)  ]^{-1}\Phi^\perp (z),
\ee
$(z\in x)$ where $\Phi^\perp $ is the component of $\Phi $ perpendicular
 to $\Psi$, the \hfill \\
 \mbox{$[\cos \theta (z)  ]^{-1}\Phi (x)\Lap \Psi $-term}
  splits into
a $\Psi \Lap \Psi$-term and a remainder which is small as a consequence of the
maximizing condition on $\Psi$. It contains no $\ln \tilde{a}/a $ piece and 
is of higher order in $\hat Q-1$.  As a result 

\ba
\Seff &=& - \sum_{x} \ln J_0( \Cav \Psi (x))
-\frac 12 Tr \ln \Gamma_Q
+ \frac 12  \int_z \beff (z)|\nabla_\mu \Psi(z)|^2 
\nn  \\
& &
+ \mbox{renormalized 1-loop diagram}
+\mbox{lattice artifacts} \ . \label{approxResult} 
\ea
The sum of the first two terms will be called the {\em augmented jacobian},
the jacobian proper is given by eq.(\ref{jacIntro}). 
The renormalized 1-loop diagram is the same as $\Seff^{(2)}$ above,
 except that 
$\Gamma_{KG}$ is substituted for $\Gamma_Q$; moreover the last term in the 
definition (\ref{alphaDef}) of $j_\mu $ may be dropped
in $\Seff^{(2)}$ because it is of order 
$a$. The effective coupling constant is
\be
\label{beff}
\beff (z) = -({\mathcal{N}-1})\Gamma_{KG}(z,z).
\ee
$\Gamma_{KG}(z,z)$ is very nearly constant except near block boundaries. 
Therefore we expect that the deviation of $\beff (z)$ 
from its block average can be neglected. Finally
\ba
\mbox{lattice artifacts} &=&  \int_z \tr j_\mu (z) \nabla_{\mu } 
\Gamma_{KG}(z,w)|_{w=z+\hat \mu }  \\
& = & -
\frac 3 8  \int_z |\nabla_\mu \Psi |^2(z)  + O(a). \nn
\ea
as $a \mapsto 0$. 
The lattice artifacts would vanish if $\Psi^T\nabla_\mu \Psi $ were zero
as is true in the continuum.
On the lattice it is $O(a)$, but nevertheless there remains a contribution
 when $a \mapsto 0$ because of the singularity of $\nabla_\mu \Gamma_{KG}$ at 
coinciding points. It amounts to a finite subtraction from the bare 
coupling constant. 

Only the $\beff $-Term contributes to order $\ln \tilde{a}/a$, and we 
recover Polyakovs result in this approximation. 
Suppose the blockspin is reasonably smooth, so that 
 $\Psi (z) - \Phi (x) $ may be regarded as a small 
quantity, of order $\epsilon$,
and $\nabla_x \Phi (x)$ is also small, $ o(\epsilon^0)$.  

Then $\hat Q -1 $ is of order $\epsilon$. We expand $Tr \ln \Gamma_Q$ to 
order $\epsilon^2$. 

To  order $\epsilon^2$ 
the augmented jacobian comes out as the sum of 
$- \sum_{x} \ln J_0$, given by eq.(\ref{jacIntro}), and 
\ba
-\frac{1}{2} Tr \ln \Gamma_Q 
&=& \int_x \int_z \Aop_{KG}(z,x)\Cav (x,z)[\cos^2\theta(z) + \cos \theta (z) -2 ] 
 \nn \\ 
& & +   \int_{z,w}\int_{x,y} 
\Psi^\perp (z)\cdot \Psi^\perp (w) [
\Gamma_{KG}(z,w)\Cav^\dagger(w,x) u^{-1}(x,y)\Cav(y,z)\nn  \\
& & \quad \quad + \Aop_{KG}(z,x)\Cav(x,w)\Aop_{KG}(w,y)\Cav(y,z)] 
\ .
\label{augToJac}
\ea
with $\Cav^\dagger(w,x)=\Cav(x,w)= \tilde a^{-2}$ if
$w\in x$ and 0 otherwise. The jacobian proper and the first term 
of the augmentation are similar but they have a different
 ${\mathcal{N}}$-dependence. 

In these formulas, $\Aop_{KG} $ and $\Gamma_{KG} $ are the Kupiainen-Gawedzki 
interpolation operator and high frequency propagator 
for scalar fields. For later use
 we indicate their definition for 
finite $\kappa $; a limit $\kappa \mapsto \infty $
can be taken at the end of the calculation. 
Let $v_{Cb} = (-\Lap )^{-1}$. Then 
\ba 
\Aop_{KG} &=& v_{Cb}\Cav^\dagger u^{-1}\ , \nn \\
u &=& \Cav^\dagger v_{Cb} \Cav \ + \frac {1} {\kappa} , \nn \\
\Gamma_{KG}
&=&  ( -\Lap + \kappa \Cav^\dagger \Cav )^{-1}   \\
\label{KGkernels}
&=& (1 - \Aop_{KG} \Cav)v_{Cb} = v_{Cb}(1 -  \Cav^\dagger  \Aop_{KG}^\dagger)\ . \nn  
\ea
Their general properties and their Fourier representation for the special 
choice (\ref{Cav}) of the averaging operator $\Cav $  are well known
\cite{GK80,GMXP96,YX97}. In particular, the propagators have exponential 
falloff 
with decay length of order one block lattice spacing $\tilde a$, and
\ba
\Cav \Gamma_{KG} &=& \frac {1}{\kappa} \Aop_{KG}^\dagger \mapsto 0 , \\
 \Gamma_{KG} \Cav^\dagger\ &=& \frac {1}{\kappa} \Aop_{KG} \mapsto 0 . 
\\
\Cav \Aop_{KG} &=& 1 - \frac 1 \kappa u^{-1} \mapsto 1 . 
\label{KGkernelsf}
\ea
The limit values are for $\kappa \mapsto \infty$.
The Fourier components of these quantities are 
recorded in Appendix D.
\footnote{A general method for proving falloff properties
of high frequency propagators which does not need translation symmetry was
described by Balaban \cite{Bal83}. }
The coordinate space expressions 
 can be evaluated 
by fast Fourier transformation. Software to do the computation and visualize
the results has been provided by Max Griessl and Jan W\"urthner and 
can be downloaded from \cite{software}, together with some screenshots. 
A sample is shown below. 

\begin{figure}[h!]
  \begin{center}
    \leavevmode
     \psfig{file=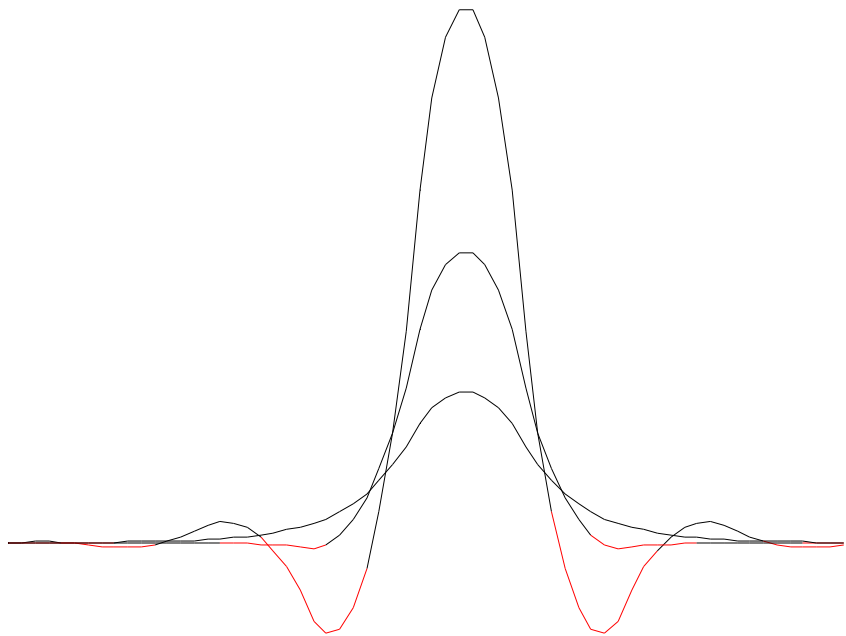, clip=, width=12cm}
  \end{center}
  \caption{\small This figure shows cross sections through A-kernels with $\kappa$ values of 100 (top), $\frac{1}{8}$ (center) and
                $\frac{1}{32}$ (bottom). Note that the kernel oscillates for high $\kappa$ values.}
  \label{kernel-cross}
\end{figure}

In principle, a calculation of the 1-loop effective action 
for pure gauge theories along the same lines is feasible. The 
linearization of the block spin constraint 
and other ingredients were described in Balaban's work
\cite{Bal85} for general gauge group. The Fourier expansions of the
interpolation kernel and high frequency propagators are also known,
for a particular choice of block spin \cite{kerres96}. 
\subsection*{Note on the large field problem}
Soon after rigorous work on the renormalization group started 20 years ago,
it was recognized that one could not expect that the effective action would 
be local for completely arbitrary block spin configurations. This was 
termed the {\em large field problem}. A device to overcome this difficulty
 was proposed by Benfatto et al \cite{Benfatto78}, and subsequently implemented
 in the
work of Kupiainen and Gawedzki \cite{GK85} and of Balaban \cite{BalII}. It involves proofs
 that {\em large fields} in the above sense are very improbable. Fermi fields
have no large field problem \cite{Feldmann86}

A {\em large field problem} 
can appear for fields which are not large in a naive 
sense. For instance, in 2-dimensional $\phi^4$-theory with a distinct 
mexican hat potential (pronounced maximum at $\phi = 0$), the block spin
$\Phi (x)\equiv 0$ is in the large field domain. Since $\Phi = 0$ is 
translation invariant, the auxiliary theory, whose field is the fluctuation 
field, ought to respect symmetry under block lattice translations, 
with symmetry group $(\tilde a{\bf Z})^2$. Numerical work by
Grie\ss l \cite{MG97}
showed that the symetry was spontaneously broken (to $(2\tilde a{\bf Z})^2$).
Such long range order in the auxiliary theory is incompatible with locality
of the effective action. 

The $\sigma$-model also has a large field problem for $\mathcal{N} > 1$.
Divide the block lattice in black and white squares in a checkerboard 
fashion and consider the configuration $\Phi (x)$ which 
points ``up'' (in +0 -direction) on white squares,
 and down on the black ones. A particular
extremizing  background field $\Psi $ has components 
\ba 
\Psi^0 (z) &=& \cos ( \pi (z^1 + z^2)/\tilde a), \nn \\
\Psi^1 (z) &=& \sin ( \pi (z^1 + z^2)/\tilde a), \nn \\
\Psi^i (z) &=& 0 \qquad (i>1) \ . 
\ea
But this is not unique. Continuous rotations in the space orthogonal to
the 0-direction yield degenerate extrema. Therefore the auxiliary theory 
has a zero mode, its correlation functions will not decay quickly, and
one cannot expect a local effective action for block spins very close to   
$\Phi $. But note that these are ``energetically'' the most unfavorable 
block spin configurations of all; they are near maxima of
 the effective action. 
\section{Linearization of the constraint}
A perturbative calculation of the functional integral (\ref{weact}) for
the
effective action is not straightforward because the argument of the
$\delta$-function is a nonlinear function of the field.

To solve this problem, we find a parametrization of an arbitrary field
$\phi $ on $\Lambda $ in terms of the background field $\Psi = \Psi [\Phi
] $
 and a fluctuation  field $\zeta $ such that the
constraint becomes a linear constraint on $\zeta $.
\be \phi (z) = \phi [\Psi , \zeta ](z) \ . \ee
The background field is a smooth field. It represents the low frequency
part
of $\phi $, while $\zeta $ adds the high frequency contributions.
$\zeta $ takes its values in a linear space.
 It has $\mathcal{N}$ components, and we choose it so that
\be \zeta (z) \perp \Phi (x) \ \mbox{for } z \in x \ . \ee
Later, a further linear transformation to variables
$\xi (z) \perp \Psi (z)$ is performed.
There is a jacobian $J$  to the transformation, and the result has the
form
\be
e^{-\Seff [\Phi ]} =   \int \prod_{z} d\zeta (z) \delta ( \Cav \zeta )
J(\Psi , \zeta )
e^{-S(\phi [\Psi , \zeta ])}. \label{Seff}
\ee
Balaban \cite{Bal85} has shown how to find a suitable
parametrization
in the case of lattice gauge fields. His method is not applicable for
the nonlinear $\sigma$-model for general $\mathcal{N}$, because it makes
essential use of the fact that the field takes values in a group, and
right and left multiplication of group elements commute,
$(g_L g)g_R = g_L(g g_R)$. But the suitable parametrization can be written
down
explicitly as follows.

Decompose $\phi (z)$ into components  $\phi^{||}(z)$ and $\phi^\perp (z)$
parallel and perpendicular to $\Phi (x) $ for $z\in x$.
The blockspin condition says that
\hfill \\  \mbox{$\Cav \phi (x)=\rho (x)^{-1}\Phi(x)$},
where $\Cav $ is a linear block average, and the scalar factor $\rho (x)$
is determined by the requirement that $\Phi (x) $ has length 1. 
If we use the block average $\Cav$ defined in eq. (\ref{Cav})
then only $\phi(z)$ with $z \in x$ enters into $\Cav \phi (x)$.
The blockspin condition is therefore equivalent to
\be
\Cav \phi^\perp (x)=0 . \label{blockLin}
\ee

We parametrize
\ba
\phi^\perp (z) &=& \Psi^\perp (z)+ \zeta (z) ,\nn  \\
\phi^{||} (z) &=& \left( 1 - \phi^\perp(z)^2 \right)^\einhalb \Phi (x)
\label{phiZeta}
\ea
for $z\in x$. Since $\Psi $ satisfies the blockspin condition
(\ref{blockLin}), the condition is equivalent to
\be \Cav \zeta  (x) = 0 \ee
The jacobian of the transformation will be worked out in Appendix A.
The result is
\ba
\label{jacobiTotal}
J(\Psi , \zeta ) &=& J^0(\Cav \phi ) \prod_{z\in x}
 \left(1 - (\Psi^\perp + \zeta)^2\right)^{-\einhalb}, \\
J^0(\Cav \phi ) &=& \prod_x J_0(\Cav \phi (x)) =
 \prod_x  (\Phi (x)\cdot  \Cav \phi (x))^{\mathcal{N}}
\label{jacobi}\ea
As usual, $\Phi $ is regarded as determined by $\Psi $.
 In a 1-loop calculation, $J_0(\Cav \phi (x))$ is approximated
 by $J_0(\Cav \Psi (x))$. 
The last factor in $J$ will be cancelled when we transform to the
$\xi $-variables.

\section{The 1-loop approximation}
\label{sec:1loop}
The 1-loop approximation yields the effective action  to order $\beta^0$.
It is obtained by expanding the action to second order and the Jacobian to
 zeroth order  in the fluctuation field. This approximates expression
(\ref{Seff}) by a Gaussian integral. The resulting $Tr \ log $ formula is
not particularly useful, though.

It is possible to obtain a first simplification by exploiting the fact
that
the background field $\Psi $ is smooth.
This is always true, whether the block spin
$\Phi $ is smooth or not, provided the blocks are chosen large enough. A
basic
reason for this is that there are no domain walls in a 2-dimensional
 ferromagnet with continuous symmetry, because the free energy of such
domain walls would decrease by making them wider. This is an old argument
by
M. Fisher \cite{MF67} which was made mathematically precise by Dobrushin and
Shlosman's in their
version of the  proof of the absence of spontaneous
breaking of continuous symmetries in 2 dimensions  \cite{DS75}.

Because of the smoothness of $\Psi $ one can neglect
terms of higher order than second in $\nabla \Psi $.
Note however that this smoothness argument {\em cannot } be used to argue
 that $\cos \theta (z)=\Psi^T (z) \Phi (z) $  must alway be close to $1$.
 Only for sufficiently {\em smooth} block spin
field $\Phi (x)$ will it be true that the component $\Psi^\perp (z)$
of $\Psi (z)  $ which is perpendicular to $\Phi (x)$ is small.

The action $S$ involves derivatives of $\phi $ which contribute
derivatives
 in $\zeta$. Because of the constraint $\zeta (z)\Phi (x) = 0$ for $z\in
x$,
$\zeta $ will have jumps at block boundaries which contribute to the
derivatives. In order to avoid this complication, it is convenient to make
a
further linear transformation from $\zeta (z) \perp \Phi (x) $ to
variables
 $\xi (z) \perp \Psi (z)$,
\ba
\xi (z) &=& Q^{-1}(z)  \zeta (z) \ , \\
 \xi (z) \Psi (z) &=& 0 \ . \label{xiConstraint}
\ea
$Q(z)$ depends on $\Psi (z)$ and $\Phi (x)$. It is a linear transformation
between different tangent spaces of the sphere
\be Q(z) : T_{\Psi (z)}\mathcal{S}^{\mathcal{N}} \mapsto
 T_{\Phi (x)}\mathcal{S}^{\mathcal{N}}\ \mbox{for } z\in x .
\label{Qmap}\ee
The tangent spaces are $\mathcal{N}$-dimensional.
We introduce the abbreviation
\be \pi(z) = 1 - \Psi(z)\Psi^T(z) . \label{pi}\ee
 In covariant form, $Q$ is as follows
\ba
Q &= & \pi + \Phi (\Psi^T \cos \theta - \Phi^T ) , \label{Qdef} \\
Q^{-1} &=& 1 - \frac{1}{\cos \theta}\Phi  \Psi^T .
\ea
An expression in a particular basis will be given in appendix A.1.
 It shows that
the modulus of the determinant of the resulting $\mathcal{N}\times \mathcal{N}$
matrix $Q(z) $ is
\be
\label{detQ} 
|\mbox{det} Q(z)| = \cos \theta (z) = (1 - (\Psi^\perp )^2 )^\einhalb .
\ee
Later on we shall introduce an extension of $Q$ to a map 
$\hat Q: \RNP \mapsto \RNP $. 

The expansion of the field $\phi (z) $ in powers of $\xi $ comes out as
\be
\phi (z) = \Psi (z)  +  \xi (z) - \frac 1{2\cos \theta (z)} \xi (z)^2 \Phi
(x)
+ \dots \label{xiParametriz}
\ee
The action $S(\phi )$ can now be expanded up to second orders in $\xi $,
\be
S(\phi ) = S(\Psi ) + \frac 12 \xi S^{\prime \prime } (\Psi) \xi + 
\mbox{linear in $\xi $ } + \dots . 
\ee
Now we are ready to consider the effective Boltzmann factor.
In one loop approximation, i.e. to order $\beta^0$
the jacobian factor gets expanded to 0-th order in the fluctuation field. 
Furthermore
$$ d\zeta (z) = |\mbox{det} Q(z)| d\xi (z) $$
and $\mbox{det} Q $ is as given above in eq.(\ref{detQ}). Therefore the
factor multiplying $J^0 $ in the jacobian in (\ref{jacobi}) cancels out
and we
 get to 1-loop order
\be
 e^{-\Seff [\Phi ]} =
 e^{-S(\Psi ) } J^0(\Cav\Psi )  \int \prod_{z} d\xi (z) \delta ( \Cav Q\xi )
e^{-\int_z \frac{\beta}{2} \left( \xi S^{\prime \prime}(\Psi )
\xi \right) } \label{Seffloop} 
\ee
There is no linear term in $\xi $ in the exponent because $\Psi $
 extremizes the action subject to the condition of fixed blockspin, and
because $\xi $ parametrizes fields with the same blockspin.

The integration of the variable $\xi (z)$ is over the
$\mathcal{N}$-dimentional  tangent space $\TSz $, i.e. subject to the 
constraint 
\be \Psi (z)\cdot \xi (z) = 0 . \label{constrPsiXi}\ee
The $\delta$-function $\delta ( \Cav Q\xi )$
 can be regarded as limit of a Gaussian. So we have
to evaluate a Gaussian integral. As a result, one obtains the effective 
action  as a sum of the classical action (tree perfect action) $S(\Psi )$, 
the jacobian term $-\ln J_0$  
and a $ Tr \ln \Gamma$-term. The propagator $\Gamma $ is the covariance
of the above mentioned Gaussian measure.

This formula is not particularly illuminating because the
full  propagator $\Gamma $
has a complicated dependence on the field $\Psi $. It comes from three sources:
The constraint (\ref{constrPsiXi}) on $\xi $, the $\Psi$-dependence of $Q$, 
and finally the $\Psi$-dependen\-ce of $S^{\prime \prime }$.

A simplification is possible because the smoothness of the background field 
$\Psi $ (on lenght scale $a=$ lattice spacing of the fine lattice)
 can be exploited. 
In the approximation which exploits the
smoothness of the background field $\Psi $ it is not necessary to consider
terms of higher than second order  in $\nabla_\mu \Psi $.
$S^{\prime \prime}$ contains field dependent terms of first and second order 
in $\nabla\Psi$. They 
can be treated as  perturbations which are  treated by second and 
first order perturbation theory, respectively.
 This extracts the field dependence of  $S^{\prime \prime}$
from the propagator. 

The field dependence in $Q$ reflects the detailed choice of the block spin.
Its contributions are not of order $\ln  \tilde{a} /a $ and are therefore not 
included in Polyakovs result. The derivation of explicit formulas 
depends on the assumption that the block spin field $\Phi $ on the
 block lattice is smooth enough, or, more precisely,
 on sufficient smoothness of 
$\Psi $ on the lenght scale of the lattice spacing $ \tilde{a} $ of the
 block lattice. 
There exists an extension of $Q$ to an 
$(\mathcal{N}+1)\times (\mathcal{N}+1)$ matrix   $\hat Q$. 
When the assumption holds, $\hat Q$  is 
close to 1, and one can derive a power series expansion of the propagator
in powers of $\hat Q -1 $. We will later compute this expansion.

There remains the constraint $\Psi \cdot \xi =0 $ on the integration variables
$\xi (z)$. There are two ways to handle this
\begin{enumerate}
\item {\em Polyakovs method}. One expands $\xi (z)$ in a basis 
$e_1(z),\dots e_{\mathcal{N}}$ for the tangent space $\TSz $. In differential 
geometry, such a basis  is called a {\em moving frame}. $S^{\prime \prime }$
becomes a $\mathcal{N}\times \mathcal{N}$ matrix in this basis. 

Polyakovs method has the advantage that the origin of the characteristic 
factor $\mathcal{N}-1$ in the formula for the running coupling constant 
emerges in a very transparent fashion from the form of $S^{\prime \prime}$.
Therefore we show the details in the next section. The result  agrees
 with Polyakovs', to order $\ln  \tilde{a} /a $.

The disadvantage of Polyakov's method is that the expansion of the propagator
in powers of $\hat Q -1$ would be very thorny.
\item {\em $\mathcal{N}+1$-dimensional integration}. Here one inserts 
$1$ in the form of a Gaussian integral over an additional integration
variable $\xi^0 (z) \in {\bf R}$. This is combined with the
 $\mathcal{N}$-dimensional  integration over $\xi (z)$ to an
 $\mathcal{N}+1$-dimensional integration over 
$\varphi (z) = \xi (z) + \xi^0 \Psi (z)\in \RNP$. 

In this formulation the power series expansion in $\hat Q -1 $ is
 straightforward, but Polyakovs result  must be 
extracted by evaluating the singular part of a 1-loop graph. 
\end{enumerate}
It is convenient to write the action in a gauge covariant form by introducing 
an arbitrary z-dependent basis. This yields results which can be used in both 
methods. The basis  
consists of an orthonormal set of vectors
 $e_\alpha (z) $, $\alpha = 0,\dots \mathcal{N}$
for every site $z\in \Lambda $ which span $\RNP$, so that
\be
e_\alpha \cdot e_\beta (z) = \delta_{\alpha \beta }.
\ee
The field can be expanded in the basis
\be
\phi (z) = \sum_{\alpha=0}^\mathcal{N} \phi^\alpha (z)e_\alpha (z) .
\ee
and similarly for $\xi $ and $\Phi $. We assemble the expansion coefficients 
in $\mathcal{N}+1$ dimensional column vectors $\phI, \PsI , \PhI $ and $\xI$. 

 One introduces
matrices $A_\mu (z) $ by
\be A_{\alpha \beta \mu }(z) = e_{\alpha }(z+\hat \mu )
         \cdot \nabla_\mu e_\beta (z)\ee
On the lattice, the Leibniz rule takes the form 
\be
\nabla_\mu (f(z)g(z)) =  (\nabla_\mu f(z))g(z)+
f(z+\hat \mu )\nabla_\mu g(z). \label{Leibniz}
\ee
Using this
one finds the following substitute for antisymmetry in indices 
$\alpha , \beta $,
\be A_{\alpha \beta \mu } (z) =A_{\beta \alpha  -\mu }(z+\hat \mu).
\label{antisymmetry}
\ee
The action takes the covariant form 
\be|\nabla_\mu \phi |^2 = |(\nabla_\mu  + A_\mu )\phI|^2 . \ee
In a constant basis, one has $A_\mu = 0$. 

The expansion (\ref{xiParametriz}) carries over to the column vectors. Using 
it one computes with the help of the lattice Leibniz rule 
\be
|\nabla \phI|^2(z) = |\nabla \PsI|^2(z) + |\nabla \xI |^2(z) +
\frac {1}{\cos \theta (z)} \PhI(x)\cdot \Lap \PsI (z)\xI^2(z)
\label{actionPre}
\ee
with $\cos \theta (z) = \Psi (z)\cdot \Phi (x)$ as usual, $z\in x$. A 
total divergence has been omitted which arises from partial integration of
a $\nabla \xI^2$-term.  

Because of the smoothness of $\Psi $, $\PhI\cdot \Lap \PsI$ is of order 
$|\nabla \Psi|^2$. To order $|\nabla \Psi|^2$ we find 
\be
\xi S^{\prime \prime}\xi = \int_z \left[ |(\nabla_\mu  + A_\mu )\xI |^2(z)
+ \frac {1}{\cos \theta (z)} \PhI (x) \cdot \Lap \PsI (z)\xI^2(z)\right] .
\label{actionCov}
\ee

\section{ Polyakov's method}
In Polyakovs method one uses a basis with 
\be
e_0 (z) = \Psi (z)\ .
\ee
The basis vectors $e_1(z), \dots e_{\mathcal{N}} (z)$ span the 
tangent space $\TSz $ and the $\xi$-field has no $0$-component. 
 
There is a remaining arbitrariness in the  choice of basis.
 The $O(\mathcal{N})$-group of those local  rotations which
leave $\Psi (z)$ invariant form a symmetry group of gauge transformations.
The $\mathcal{N}\times \mathcal{N}$ matrices
\be a_\mu (z) = (A_{ij\mu}(z), \ i,j=1\dots \mathcal{N}) \ee
transform like gauge fields under these gauge transformations, while
\be A_{i0\mu }(z) = \left( \nabla_\mu \Psi (z)  \right)^i \ee
transform like $\mathcal{N}$-vector fields.

We compute the field strength tensor
\footnote{This is the field strength tensor 
which one gets by use  of
noncommutative differential calculus. It was shown by Dimakis,
 M\"uller-Hoissen and Striker \cite{MuellerHoissen}
that the conventional lattice gauge theory formalism is equivalent to a
 noncommutative differential geometry.
 In this formulation, the lattice Leibniz rule 
eq.(\ref{Leibniz}) above takes the standard form $ d(fg)= (df)g + f dg$,
and all the familiar formula of continuum gauge field theory remain valid 
on the lattice. } 
for the vector potential $a_{\mu}$,
\ba
F_{ij\mu \nu } (z) &=& \namu a_{ij\nu}(z)-\nanu
a_{ij\mu}(z) \nn
\\ & & 
+a_{ik\mu}(z+\nu)a_{kj\nu}(z)-a_{ik\nu}(z+\mu)a_{kj\mu}(z) \ .
\label{Fmunu}
\ea
Using the completeness relation for the basis in the form
$$\sum_{i=1}^{\mathcal{N}} e_i(z) e_i(z)^T
= 1 - \Psi \Psi^T $$
one computes the component of the field strength tensor as
\be
F_{ij\mu\nu} = e_i(z+\hat \nu + \hat \mu )\nabla_\mu \Psi (z+\hat \nu)
e_j(z) \cdot  \nabla_\mu \Psi (z) - (\mu \leftrightarrow \nu ) .
\ee
We see that the field strength tensor  is of order $(\nabla \Psi )^2 $.
 It follows that the vector potential $a_\mu $ in Lorentz gauge,
$$ \nabla_\mu a_\mu = 0 \ , $$
is also of order $(\nabla \Psi )^2 $. The $[a_\mu , a_\nu ]$ term in
expression (\ref{Fmunu}) is negligible and the vector potential to
 leading order could be recovered as
$$ a_\nu (z) = - \int_w \nabla_\mu v_{\mathrm{Cb}} (z-w) F_{\nu \mu }(w) \ .  $$
Although the Coulomb potential $v_{\mathrm{Cb}} $ in 2 dimensions does not
exist, its derivative is well defined.

Separating the terms which involve $a_\mu $ and $A_{0i\mu} $
and using the antisymmetry eq.(\ref{antisymmetry}) of $A$ and 
$\nabla_{-\mu }\Psi (z + \hat \mu )= -\nabla_\mu \Psi (z)$  we obtain
\be
|(\nabla_\mu + A_\mu )\xI|^2 = |(\nabla_\mu  + a_\mu )\xI|^2 +
|\xi \cdot \nabla_\mu \Psi |^2 . 
\ee
The last term involves the components $[\nabla \Psi]^i $ of $\nabla \Psi $ 
with respect to the moving frame,
 {\em not} $\nabla \Psi^i $.

In conclusion
\be
\xi S^{\prime \prime }\xi= \int_z |(\nabla_\mu + a_\mu )\xI |^2(z)  + 
\xi S_I^{\prime \prime}  \xi \label{S''}
\ee
with
\be
\xi S_I^{\prime \prime}  \xi = \int_z \left(
[\xi \cdot \nabla_\mu \Psi ]^2(z) + \xi^2(z)[\Psi \cdot \Lap \Psi ](z) 
+ \frac {1}{\cos \theta } \xi^2 (z) [\Phi^\perp \cdot \Lap \Psi ] (z)\right). \nn
\ee

The $\Phi \cdot \Lap \Psi $-term was split into two terms in order to 
single out the last term in $S_I^{\prime \prime}$.
We will see later that this last term is very small for
smooth enough block spin fields.
This is a consequence of the extremizing property of $\Psi $.
The term is $\Lap \Psi $ multiplied with an expression of order
 $(\Psi^\perp)^3 $, and turns out not to contribute at all to order 
$\ln \tilde{a} /a $. 

The effective Boltzmann factor becomes
\be
 e^{-\Seff [\Phi ]} =
 e^{-S(\Psi ) }  \int \prod_{z} d\xi (z) \delta ( \Cav Q\xi )
J^0(\Cav\Psi )
e^{-\int_z \frac{\beta}{2} \left([ (\nabla_\mu + a_\mu) \xi ]^2 + \xi
S_I^{\prime \prime}(\Psi )
\xi \right) } \label{Seff1loop}
\ee
The $\delta$-function can be regarded as limit of a Gaussian, and we have
to evaluate a Gaussian integral.

Let us write $[z]\in \tilde{\Lambda} $ for the block $x$ which contains
$z$.
Let us remember that $Q(z)$ is a map (\ref{Qmap}) from
$ T_{\Psi (z)}\mathcal{S}^{\mathcal{N}}$ to
$ T_{\Phi (x)}\mathcal{S}^{\mathcal{N}}$, and $\Cav $ defines a map of
 functions on the fine lattice $\Lambda$ with values in $T_{\Phi ([z])} \mathcal{S}^{\mathcal{N}}$
 to functions on the block lattice with values in $T_{\Phi (x)} \mathcal{S}^{\mathcal{N}}$.
Therefore
the operator $Q^T \Cav^\dagger \Cav Q $  maps functions with values in
 $ T_{\Psi (z)}\mathcal{S}^{\mathcal{N}}$ into functions of the same kind.
The Polyakov basis elements $e_i(z), \ i=1\dots \mathcal{N}$ are a basis
for
 $ T_{\Psi (z)}\mathcal{S}^{\mathcal{N}}$.
We denote by $P(z,w)=(P_{ij}(z,w), ij=1\dots \mathcal{N}) $
 the matrix of the kernel of
$Q^T \Cav^\dagger \Cav Q $ with respect to this Polyakov basis, viz.
\be \int_w P_{ij}(z,w)\xi^j(w)  = e_i(z) \cdot
\left( Q^T \Cav^\dagger \Cav Q \xi \right) (z).
\ee
$P (z,w) $ is only nonzero when $z$ and $w$ belong to the same block $x$.
The $\delta$-function becomes the limit of a Gaussian as follows
\ba
\delta ( \Cav Q\xi )  &=& \lim_{\kappa \mapsto \infty } N_\kappa
e^{-\frac {\beta \kappa} 2 \int_z \int_w  \xi^i(z)P_{ij}(z,w)\xi^j (w)}\\
N_\kappa &=& \left( {\kappa a^d}/{2\pi}  \right)^{\mathcal{N}/2}
\ea
Define the high frequency propagator  (=propagator of the $\xi$-field)
in the Polyakov basis
\be
\Gamma^e_\kappa  = \beta^{-1} \left( - [\nabla_\mu + a_\mu ]^2 + \kappa P
\right)^{-1} \label{Gamma_e}
\ee
Now we can evaluate expression (\ref{Seff1loop}) for $\Seff $
with volume element 
$$ d\xi (z) = d\xi^1(z)\dots d\xi^{\mathcal{N}}(z) . $$
The result is 
\be
\Seff[\Phi ] = -\ln J^0(\Cav \Psi )+  S (\Psi ) +\frac{1}{2} \tr S_I^{\prime
\prime }(\Psi )\Gamma^e_{\kappa }
 -\frac 12 Tr \ln \Gamma^e_{\kappa }   + const
\ee
in the limit $\kappa \mapsto \infty$.
Note that $P$ depends on $\Psi $ because $Q$ depends on $\Psi  $.
\footnote{in addition there is an implicit $\Psi$-dependence through 
the moving frame and through $a_\mu$.}
Therefore
the propagator $\Gamma^e_{\kappa } $ also has a residual $\Psi $
dependence.
It is small when the block spin field is smooth, because 
the extension $\hat Q$ of $Q$ is in this
case
close to $1$. Unfortunately it would be difficult to find the first order term 
in $\hat Q -1 $ in this formalism, because the formula for $P$ contains the 
moving frame, and because there could be a term which is first order both in 
$a_\mu $ and in $\hat Q -1$.  

\subsection{Recovery of Polyakov's result}
\label{PolyakovResult}
Polyakov determined the contributions to the effective action
 which are of order $\ln \tilde{a}/a$. They do not
depend on the detailed form of the blockspin 
which fixes the infrared cutoff in the auxiliary theory with fields $\xi$.
The term $\kappa P $ in the high frequency propagator has the  effect of
an
 infrared cutoff. This has been discussed in detail in the work of
Kupiainen and Gawedzki \cite{GK80}. To get the result modulo details of the
choice of infrared cutoff,
 we may therefore replace $\kappa P$ by a mass term $M^2
$,
where $M= o(\tilde{a}^{-1})$.

The propagator also has a dependence on the $O(\mathcal{N})$-gauge
 field $a_\mu $. We show that this can be neglected, by exploiting the
smoothness of the background field $\Psi$. We need only consider terms up
to order
 $O([\nabla \Psi]^2).$
The result is gauge invariant.
 $a_\mu$ in Lorentz gauge is $O([\nabla \Psi]^2)$ as we saw.
 A perturbation expansion in $a_\mu $ shows that $Tr \ln \Gamma^e
  = O(a_\mu^2)$. Therefore the $a_\mu $ dependence of this term can
 be neglected. $S_I $ is already $O([\nabla \Psi]^2)$, therefore the
 $a_\mu$-dependence in the propagator multiplying it can also be
neglected.
The high frequency propagator matrix can therefore be replaced by
\be \Gamma^e_\kappa (z,w)_{ij} \approx \beta^{-1}\delta_{ij}v_M(z-w) \ee
where $v_M$ is the Yukawa potential in 2 dimensions with mass $M$ 
of order $\tilde{a}^{-1}$, viz.
 $$v_M = (-\Lap +M^2)^{-1} $$
The $Tr \ln $-term has become a constant.
The jacobian is not ultraviolet divergent and is therefore a feature of
the
details of the infrared cutoff.
Inserting $S_I^{\prime\prime} $ we get the result in the desired
approximation
\be
\Seff [\Phi ] = \frac 12\int_z (\beta - (\mathcal{N}-1)v_M(0))
[\nabla_\mu \Psi (z)]^2  + S_{\mathrm{isZero}} ,
\label{polyakovResult}
\ee
with
\be
S_{\mathrm{isZero}} = \frac 12 \mathcal{N} v_M(0)\int_z [\cos \theta (z)]^{-1}
 \Phi^\perp \cdot \Lap \Psi (z)
\ee
Here as everywhere
\ba \cos \theta (z) &=& \Phi (x)\cdot \Psi(z), \\
\Phi^\perp (z) &=& \Phi (x) - \Psi (z) (\Phi (x)\cdot \Psi(z))
\ea
for $z\in x$. $\Phi^\perp $ is the component of the blockspin which is
perpendicular to the background field.

Except for the  term $S_{\mathrm{isZero}}$ this is Polyakov's result.
We show in section \ref{sec:polyakovCorr} that $S_{\mathrm{isZero}}$
is actually zero
as a consequence of the extremality condition on the background field $\Psi$.

Thus, Polyakov's result has been recovered.
%
\subsection{A note on high frequency propagators}
\label{sec:NoteOnHF}

We record here a formula for the {\em full} high frequency propagator 
$\Gamma $ which would figure in the ``not very illuminating'' formula
\be \Seff [\Phi ] = - \ln J^0(\Cav \Psi) + S(\Psi) - \frac 12 Tr \ln \Gamma 
\ee 
as mentioned earlier. It is obtained 
by inspection of the exponent in the integral representation (\ref{Seff1loop}),
the alternative fromula
(\ref{GammaAlt}) is obtained from the alternative treatment using
the constant basis in section \ref{sec:NDim} below in the same way. 
\ba
\Gamma (z,w) &=& \sum_{i=1}^{\mathcal{N}} e_i^T(z)
([\nabla_\mu + a_\mu ][\nabla_{-\mu} + a_{-\mu}] \label{Gamma}   \\
  & &\quad + \nabla_\mu\Psi \nabla_\mu \Psi^T 
+ [\Psi \cdot \Delta \Psi + (\cos \theta)^{-1} \Phi^\perp \cdot \Delta \Psi ]{\bf 1}
+ \kappa P)^{-1} e_i(w) 
 \\
&=& \pi (z)(-\Lap + 2 \nabla_\mu \Psi \nabla_\mu \Psi^T + 
\Psi \nabla_\mu\Psi^T \nabla_\mu \Psi \Psi^T  
+ \leftNabla j_\mu + j_\mu^T \nabla_\mu  \nn \\
& & \qquad \qquad + 
+ (\cos \theta)^{-1} \Phi \cdot \Delta \Psi  \pi +
\kappa \hat Q^T \Cav^\dagger \Cav \hat Q  )^{-1}\pi (w)\ ,
\label{GammaAlt}
\ea 
with the understanding that $e^T_i$ are the basis vectors in the dual space,
and
\be
(j_\mu \varphi)(z) = 
j_\mu (z)\varphi(z+\hat \mu ), 
\ee
i.e. $j_\mu $ contains a shift operator. 
(Remember the footnote on noncommutative differential calculus.). 
Here as everywhere, $\pi (z) = 1 - \Psi \Psi^T $ projects on 
$\TSz$. We see from the second formulae that
\be
\Gamma = \pi \Gamma_{Q[\Psi]}\pi + O(\nabla \Psi).  \label{GammaAlt1}
\ee
There is a correction term of {\em first order} in $\nabla \Psi$ because
$j_\mu $ is of first order in $\nabla \Psi$, see eq.(\ref{alphaDef}). 
This explains why $Tr \ln \Gamma $ produces among others a 
 1-loop graph (\ref{ren1loop})
which involves $\Psi\nabla\Psi^T (z)$ at two different sites. 
\section{$\mathcal{N}+1$-dimensional integration}
\label{sec:NDim}
We present now the alternative method for evaluating the Gaussian integral 
(\ref{Seff1loop}) for the effective action. This will prepare the ground for
 the expansion of the result in powers of $\hat Q -1$.

We insert extra integration variables $\xi^0(z) \in {\bf R}$ by insertion of 
\be
1 = N_\kappa \int \prod d\xi^0 
e^{- \frac{\beta}{2} \left( \int_z |\nabla \xi^0 |^2 + \kappa \xi^0 \Cav^\dagger \Cav \xi^0
\right)}\ee 
$N_\kappa $ is a constant which is not field dependent, and $\Cav $ is the
block average similarly as before. 

We will combine the integration variables $\xi^0 $ and $\xi $ to 
\be
\varphi (z) = \xi (z) + \xi^0(z)\Psi (z) 
\ee 
so that 
\ba
\xi^0 &=& \Psi^T \varphi  \nn \\
\xi &=& \pi \varphi , \nn  \\
\pi (z) &=& 1 - \Psi (z) \Psi^T (z) \ . \label{ids} 
\ea
Here and in the everywhere we write superscripts $T$ for the transpose. 
The transpose $\Psi^T$ of a column vector $\Psi $ is a row vector. 
 
$\xi (z) $ is now considered as an element of $\RNP$. It satifies the 
constraint $\Psi^T \xi =0$. The symbol $\nabla_\mu \xi $ will stand 
for the finite difference derivative of this $\RNP$-valued field. 
In other words, we expand now in a constant basis
 $e_0, \dots ,e_{\mathcal{N}}$,
viz. the natural basis for $\RNP$. In this way we can use the result 
eq.(\ref{actionCov}) with $A=0$,
and we can write  $\xi $ in place of $\xI$ etc.   

Adding the $|\nabla \xi^0 |^2 $ term to the action, we obtain an 
extended action
\ba 
S_{\mathrm{ext}}(\Psi | \varphi ) &=& \frac {\beta}{2}\int |\nabla \xi^0|^2 +
  S(\phi )\nonumber
\\
&=& S(\Psi) + \frac {\beta}{2}\int_z \left( |\nabla\xI |^2 +
|\nabla \xi^0|^2 
+ \frac {1}{\cos \theta (z)} \PhI (x) \cdot \Lap \PsI (z)\xI^2 
\right)\nonumber
\\
&\equiv & S(\Psi ) +  \frac {\beta }{2}   \int_z \left( |\nabla_\mu \varphi |^2 (z)
+
\varphi^T  S_{\mathrm{ext,I}}^{\prime \prime } \varphi + \dots \right)
\ea
In Appendix B the sum of the first terms is computed. As a 
result
\ba
\varphi^T  S_{\mathrm{ext,I}}^{\prime \prime } \varphi &=& 
\varphi^T(z+\hat \mu)\left[ 2 \nabla_\mu \Psi \nabla_\mu \Psi^T    
+ \Psi (z+\hat \mu )\nabla_\mu \Psi^T \nabla_\mu \Psi \Psi^T (z +\hat \mu ) \right]
\varphi(z +\hat \mu ) \nn
\\
& & +  \frac {\varphi^T\pi\varphi}{\cos \theta (z)} \Phi^T (x) \Lap \Psi (z) 
+ (\nabla_\mu \varphi^T j_\mu \varphi (z+\hat \mu ) + \mbox{transpose}) \ . 
 \nn
\ea
Repeated indices $\mu $ are summed over. 
The  $\delta $-function $\delta (\Cav Q\xi )$ is again considered as a limit
of a Gaussian. Its exponent combines with 
$  \beta \kappa \xi^0 \Cav^\dagger \Cav \xi^0/2 $ according to 
\be 
 \varphi^T \hat Q^T \Cav^\dagger \Cav \hat Q \varphi =  
\xi^0 \Cav^\dagger \Cav \xi^0 + \xi^T Q^T \Cav^\dagger \Cav Q \xi   \label{QhatC}
\ee
with $\hat Q$ as follows. 

The definition (\ref{Qdef}) of $Q$ extends to a map 
$\RNP \mapsto \RNP $ which has the property that it annihilates $\Psi (z) $ and 
maps $\TSz $ to $\TSx \subset \RNP $. We add to this the operator 
$\Phi(x) \Psi^T (z)  $ which annihilates $\TSz $ and maps the ray through
 $\Psi (z) $ into the ray through $\Phi (x)$. This gives 
\be 
\hat Q(z)  = 1 - \Psi (z)\Psi^T (z) + 
\Phi (x) (\Psi^T (z)[1+\cos \theta (z)] - \Phi^T (x)). \label{Qhat}
\ee
Using the indicated ranges of the various maps
and eqs.(\ref{ids}), it is readily verified that 
formula (\ref{QhatC}) holds true. 

Now we are ready to evaluate the Gaussian integral which defines the effective action
\be
 e^{-\Seff [\Phi ]} =
 e^{-S(\Psi ) } J^0(\Cav \Psi )  \int \prod_{z} d^{\mathcal{N}+1}\varphi (z) 
e^{- \frac{\beta}{2}\int \left[ \kappa \varphi^T \hat Q^T \Cav^\dagger \Cav \hat Q \varphi + 
 |\nabla_\mu \varphi |^2  +  
 \varphi^T S_{\mathrm{ext,I}}^{\prime \prime}\varphi  \right] }
 \label{Seff1loopAlt} \\
\ee
A limit $\kappa \mapsto \infty $ is to be taken in the end.

We define the new high frequency propagator 
\be
\Gamma_Q = \left( -\Lap + \kappa \hat Q^T \Cav^\dagger \Cav \hat Q \right)^{-1}  
\label{GammaQ}
\ee
$\Gamma_Q$ depends on $\Psi $ through $\hat Q$. When we want to make this 
dependence explicit, we write $\Gamma_{Q[\Psi ]}$.

$\Gamma_Q (z,w)$ is a map $\RNP \mapsto \RNP$, i.e. an 
$\mathcal{N}+1 \times \mathcal{N}+1 $ matrix. Its only $\Psi$-dependence is 
in $\hat Q$. In section \ref{sec:fieldDep} we will show how to expand in a power series in $\hat Q -1 $. In zeroth order, $\Gamma  $ agrees with the 
Kupiainen Gawedzki high frequency propagator \cite{GK80},
\be
(\Gamma_Q)_{\alpha \beta} = \Gamma_{KG}\delta_{\alpha \beta} + O(\hat Q -1) 
\ee

Using this propagator, the effective action can be computed by perturbation
 theory. Because of the smoothness of $\Psi $, we are only interested in terms 
up to order $|\nabla \Psi|^2$. But $j_\mu $ is of first order in $\nabla\Psi$.
Therefore the $j_\mu$-term must be treated to {\em second} order,
while all the other terms need only included to first order in the perturbation expansion.
As a result 
\ba
S_\mathrm{eff}[\Phi ] 
&=& S(\Psi ) - \ln J_0(\Psi ) -\frac 12 Tr \ln \Gamma_{Q[\Psi]} 
+ \frac 12 \langle[\varphi^T S^{\prime\prime}_{\mathrm{ext,I}}(\Psi) \varphi]\rangle
\nn  \\
& &   - \frac 1 8 \langle[\varphi^T S^{\prime\prime}_{\mathrm{ext,I}}
(\Psi) \varphi]^2\rangle^T 
\ea
where $\langle \cdot \rangle$ is the expectation value in a free field theory
with propagator $\Gamma_Q$ of $\varphi$, and $\langle f^2 \rangle^T
= \langle f^2 \rangle - \langle f \rangle^2.$

The expectation values can be evaluated.
 The  correction term of second order in $j$  yields
(after a change of summation variables $z,w,\mu,\nu$)
\ba
-\frac 1 8 \langle[\dots]^2\rangle^T &=&
-\frac 12  \int_z\int_w \tr \left( \nabla_\mu \Gamma_Q(z,w) \leftNabla_\nu 
j_\nu^T(w)\Gamma_Q(w,z) j_\mu(z)  \right. \nn \\
&& \left. \qquad \qquad 
+ \nabla_\mu \Gamma_Q (z,w) j_\nu(w) \nabla_\nu \Gamma_Q (w,z) 
j_\mu(z) 
\right) + \dots \ \label{C2} 
\ea
The term is  logarithmically divergent as $a\mapsto 0 $.

The first order correction is 
\ba
\frac 12 \langle[\dots]\rangle &=& \frac 12 
\int_z \tr \left[ 2 \nabla_\mu\Psi \nabla_\mu \Psi^T +  
\Psi(z+\hat \mu ) \nabla_\mu \Psi^T \nabla_\mu \Psi \Psi^T(z+\hat \mu) \right]
\Gamma_Q(z + \hat \mu , z+\hat \mu ) \nn \\
& & + \int_z \tr j_\mu (z) \nabla_\mu \Gamma_Q(z,w)|_{w=z+\mu }
+ \tr \left(\pi \Gamma_Q(z,z)\right) \frac 1 {2 \cos \theta }
\Phi^T([z])\Delta \Psi \ ;
\ea
unwritten arguments are $z$. 
The $Tr \ln \Gamma_Q $-term is needed
in $S_\mathrm{eff}$ because of the $\Psi $ dependence of 
$\Gamma_Q $; it becomes constant in zeroth order in $\hat Q -1 $.

From this we obtain the final result (\ref{result}) by 
adding to the second order correction the 
$\delta_{\mu \nu}\delta(z-w)$-term in expression (\ref{ren1loop}),
and subtracting it from the first order term. 
We show in Appendix C that this is the
 appropriate subtraction which 
renders the 2-vertex diagram convergent in the limit $a\mapsto 0$. 
When $j_\mu $ is inserted, the subtraction from the first order term
leads to a partial cancellation. The last term in the definition
(\ref{alphaDef}) of $j_\mu$
 can be dropped in eq.(\ref{C2}) and in the subtraction because its 
contributions
will be of higher order in $\nabla \Psi$ by eq.(\ref{psiNablaPsi}) below.

\subsection{Evaluation of a lattice correction term} 
\label{sec:latticeCorrection}
In order to get the simplified result in zeroth order in $\hat Q -1$,
 we need to 
also evaluate the lattice artifacts which come from the following term in
 eq.(\ref{result})
\be
\label{spurstrom} 
\tr j_\mu (z)\nabla_\mu \Gamma_{KG}(z,w)|_{w=z+\hat \mu} =
 3 \int_z  ( \nabla_\mu \Psi^T \Psi (z)+ O(a^2)) \nabla_{\mu} \Gamma_{KG}(z,w)|_{w=z+\hat \mu} .   
\ee
This is a lattice artifact; in the continuum limit 
$\nabla_\mu \Psi^T \Psi = 0$. On the lattice it is of order $a$. Nevertheless 
it cannot be neglected because 
\ba
\nabla_\mu  \Gamma_{KG} (z,w)|_{w=z+\hat \mu}
&=&-\nabla_{-\mu }\Gamma_{KG} (z+\hat \mu,w)|_{w=z+\hat \mu}  \\
&=& - \frac a 4 \Delta \Gamma_{KG}(z+\hat \mu,w )_{ w=z+\hat \mu}
 = \frac 1 {4a} + O(1)
\ea
This holds true because the singular part of $\Gamma_{KG}$ 
is translation invariant and because $\nabla_\mu \Gamma_{KG}(z,w)|_{w=z}$
is independent of $\pm \mu $ by lattice symmetry, 
while $\sum_{\pm \mu}\nabla_\mu f = a \Delta  f $. 

On the other hand (unwritten arguments are $z$)
\be
0 = \nabla_\mu(\Psi^T \Psi )(z) = \nabla_\mu \Psi^T \Psi (z+\hat \mu ) 
+ \Psi^T \nabla_\mu \Psi 
\ee
hence 
\be
\Psi^T \nabla _\mu \Psi  = - \frac a 2 |\nabla_\mu \Psi|^2 \quad \mbox{no sum}
\label{psiNablaPsi}
\ee 
Therefore the lattice artifacts are as stated in the introduction,
\be 
\tr j_\mu (z)\nabla_\mu \Gamma_{KG}(z,w)|_{w=z+\hat \mu} =
- \frac 3 8 |\nabla_\mu \Psi|^2 (z) + O(a).
\ee 
%
\subsection{Field dependence of high frequency propagator}
\label{sec:fieldDep}
Here we consider the expansion of the high frequency propagator 
(\ref{GammaQ}) in powers of $\hat Q -1$. 

Consider a propagator of the following form which depends on a real 
parameter $\alpha $
\be
\Gamma_\alpha = 
 ( -\Lap + \kappa \Cav^\dagger_\alpha \Cav_\alpha )^{-1}  
\ee
where $\Cav_\alpha $ is a (matrix valued ) $\alpha$-dependent 
block averaging operator.
A limit $\kappa \mapsto \infty$ should be taken in the end,
if desired. In our application
\be
\Cav_\alpha = {\bf 1} \Cav  + \alpha \Cav (\hat Q -1 ). 
\ee

We will use  a formula which gives the derivative  $\Gamma^\prime_\alpha $
 of $\Gamma_\alpha $
 with respect to $\alpha $. 

Let $v=\Lap^{-1}$. 
It is known from the work of Kupiainen and Gawedzki, that $\Gamma_\alpha $
admits the following representation 
\be
\Gamma_\alpha = 
(1 - \Aop_\alpha  \Cav_\alpha )v
\ee
where $\Aop_\alpha $ is an interpolation operator which maps functions on the 
coarse lattice into smooth functions on the fine lattice, and which obeys
\be
\Cav_\alpha \Aop_\alpha = 1 - \frac 1 \kappa u_\alpha^{-1}.
\ee
For finite $\kappa $, 
\ba
\Aop_\alpha &=& v \Cav_\alpha^\dagger u_\alpha^{-1}, \nn \\
u_\alpha  &=&  \Cav_\alpha v \Cav_\alpha^\dagger + \frac 1 \kappa .
\ea 
We denote differentiation with respect to $\alpha $ by a prime. 
Since
 $(u_\alpha^{-1})^\prime = - u^{-1}_\alpha u^\prime_\alpha u^{-1}_\alpha $,
one obtains by straightfoward differentiation
\ba
 \Gamma^\prime_\alpha &=& 
-(\Gamma_\alpha \Cav^{\dagger \prime}_\alpha \Aop^\dagger_\alpha +
\Aop_\alpha \Cav^\prime_\alpha \Gamma_\alpha ) \\
\Aop_\alpha^\prime &=&
 \Gamma_\alpha \Cav_\alpha ^{\dagger \prime} u_\alpha^{-1} 
- \Aop_\alpha \Cav_\alpha^\prime \Aop_\alpha \ .
\ea
In our application,
$\Cav^\prime_\alpha = \Cav(\hat Q -1 ) $ independent of $\alpha $, and
$\Aop_{\alpha=0}$ is the Kupiainen-Gawedzki interpolation operator 
$\Aop_{KG}$ 
multiplied by the $(\mathcal{N}+1)\times (\mathcal{N}+1)$ unit matrix 
${\bf 1}$. 
Therefore the expansion of $Tr\ln\Gamma_Q $ to second order in $\hat Q -1$ reads
\be
\label{expGammaQ}
Tr \ln \Gamma_Q = Tr\ln \Gamma_{KG}
     +Tr\left(\Gamma_\alpha^\prime \Gamma_\alpha^{-1} \right)|_{\alpha=0}
  +\frac{1}{2}Tr\left(\Gamma_\alpha^\prime \Gamma_\alpha^{-1} \right)^\prime |_{\alpha=0}
                    +\dots
\ee
with
\be
Tr\left(\Gamma_\alpha^\prime \Gamma_\alpha^{-1} \right)
 =-Tr\left[\Aop_\alpha\Cav(\hat Q -1) +\mathrm{h.c.} \right]
\ee
and
\ba
Tr\left(\Gamma_\alpha^\prime \Gamma_\alpha^{-1} \right)^\prime 
  &=&-Tr\left[\Aop_\alpha^\prime\Cav_\alpha^\prime + \mathrm{h.c.}\right] \\
  &=&- Tr\left[\Gamma_\alpha (\Cav_\alpha^\dagger)^\prime u_\alpha^{-1}\Cav_\alpha^\prime
 -\Aop_\alpha\Cav_\alpha^\prime \Aop_\alpha\Cav_\alpha^\prime +\mathrm{h.c.}       \right]_{\alpha=0} \label{trln2ndOrder}
\ea
The first term in eq. (\ref{expGammaQ}) is a field independent constant.

 The kernels $\Aop_{\alpha=0} $ and $\Cav $ 
are proportional to the 
unit matrix, therefore the first order term involves 
\be 
\tr (\hat Q -1 ) = \cos^2 \theta + \cos \theta -2 . 
\ee 
As a result
\be
-\frac{1}{2}
Tr\left(\Gamma_\alpha^\prime \Gamma_\alpha^{-1} \right) =
\int_x \int_z \Aop_{KG}(z,x)\Cav (x,z) 
[\cos^2\theta(z)+\cos \theta (z) -2] 
+ \mathrm{const} \label{trlnGammaQ1}
\ee

It can now be inserted into the result for the effective action. 

It remains to examine the second order term.
To order $\epsilon $
$$ 
\hat Q -1 = \Phi (x) \Psi^T(z) - \Psi(z)\Phi^T(x) + \dots 
= -(\hat Q -1)^T + \dots
$$ 
for $z\in x$.

Since the kernels $\Aop_{\alpha=0}, \Cav , \Gamma_{KG}{\bf 1} $ 
are proportional to the 
unit matrix, the second order term is an integral whose integrand contains a 
factor
\be
\tr (\hat Q -1 )(z)(\hat Q -1 )(w) = 
2[\Phi (w)\cdot \Psi^\perp(z)\ \Phi (x)\cdot \Psi^\perp(w) 
- \Psi^\perp(w)\cdot \Psi^\perp(z) \ \Phi(z)\cdot \Phi (w) ] \nn
\ee
for $z\in x, w\in y$.
The factors $\Psi^\perp $ are of order $\epsilon$, therefore the factors
$\Phi^T (\cdot )$ are only needed to order $\epsilon^0$. 
Because of the falloff properties of the kernels, 
$x$ and $y$ are either the same
or nearby blocks. Therefore if $\nabla_x \Phi (x) = o(\epsilon)$, 
we may approximate
\ba
\tr (\hat Q -1 )(z)(\hat Q -1 )(w) &=& -2 \Psi^\perp(z)\cdot \Psi^\perp (w) \ .
\ea
This may now be inserted into eq.(\ref{trln2ndOrder}) to yield the result 
for the second order contribution
\ba
-\frac 1 4 
Tr\left(\Gamma_\alpha^\prime \Gamma_\alpha^{-1} \right)^\prime_{\alpha=0} 
&=&    \int_{z,w}\int_{x,y} 
\Psi^\perp (z)\cdot \Psi^\perp (w) [
\Gamma_{KG}(z,w)\Cav^\dagger(w,x) u^{-1}(x,y)\Cav(y,z)\nn  \\
& &\quad \quad + \Aop_{KG}(z,x)\Cav(x,w)\Aop(w,y)\Cav(y,z)] 
\ .
\ea
These results are also valid for finite $\kappa $. 

Summing the two terms we obtain the result eq.(\ref{augToJac}) for 
the augmentation of the jacobian. 




\section{The background field}
\label{sec:background}
Given the block spin field $\Phi $ on the coarse lattice,
we seek the field $\phi =\Psi $ on the fine lattice which extremizes  the 
action $S(\phi )$ of the $\sigma$-model subject to the constraint 
\be C\Psi = \Phi \label{block} \ee
 The extremality condition leads to a nonlinear equation for 
$\Psi $ (eq.(\ref{eqPsi}) below). 
It is nonlinear because $\Psi (z) $ must have length 1. 

Our strategy is to start with an approximation
 $\Psi^{(0)}$ which satisfies 
the block spin condition exactly, which has the expected smoothness 
properties of $\Psi $ except for discontinuities of the 
normal derivatives at block boundaries which are small if
 $\Phi $ is resonably smooth, and 
which reduces to the exact extremum when $\Phi $ is constant.
 Starting from $\Psi^{(0)}$ we can derive improved 
approximations $\Psi^{(k)}$, $k=1,2,... $ by iteration. 
 We will see that the smoothness of $\Psi $ can again 
be exploited to argue that a single iteration with result
 $\Psi^{(1)}$ is enough is $\Phi $ is reasonably smooth. 
The formula for $\Psi^{(1)} $ will involve the high frequency propagator 
$\Gamma_{Q[\Psi^{(0)}]}$
which was encountered before. This propagator contains a dependence on
 $\Psi^{(0)}$ through $Q$. 

The formula for $\Psi^{(0)} $ will be derived in subsection \ref{subsec:psi0} below. 

We proceed to the iteration step.     Given any approximate extremum
$\Psi^{(0)}$, we parametrize an arbitrary field $\phi $ with the desired 
block spin with $\xi$-variables $\xi \perp \Psi^{(0)}$ similarly as before in 
eq.(\ref{xiParametriz}), except that $\Psi^{(0)}$ is substituted for $\Psi $.
 To first order in $\xi $,
\be
\Psi = \Psi^{(0)}+ \xi  \label{lin}
\ee 
The $\xi$-field must satisfy the constraint
\be
\Cav Q[\Psi^{(0)}]\xi = 0 . \label{xiConstr0}
\ee
The extremality condition reads 
\be
\xi^T S^\prime (\Psi ) = 0
\ee
for arbitrary $\xi \perp \Psi^{(0)}$ which satisfies the constraint  
(\ref{xiConstr0}). This is equivalent to 
\be
 S^\prime (\Psi ) = Q^T \Cav^\dagger \lambda \label{extr1}
\ee
with a Lagrange multiplier $\lambda$ which is a field on the coarse lattice. 
Power series expansion to first order around $\Psi^{0)}$ gives
$$
 S^\prime (\Psi )= S^\prime (\Psi^{(0)})+ S^{\prime\prime}(\Psi^{(0)})\xi + \dots
$$
Using eq.(\ref{lin}) one computes 
$S^\prime (\Psi^{(0)}(z)) = -\Lap \Psi^{(0)} (z) $.
But this is only valid as a linear form on 
the tangent space $T_\Psi^{(0)}S^{\mathcal{N}}$,
i.e. when contracted wit arbitrary $\xi(z)\perp \Psi^{(0)}(z)$.
 In order to remember this fact it is better to write the formula as 
\be
 S^\prime (\Psi^{(0)}(z) = -\pi^{(0)}(z)\Lap \Psi^{(0)} (z),
\ee  
with the projector $\pi^{(0)} $ on the tangent space.
 Inserting everything into eq.(\ref{extr1}) we get a linear equation for $\xi$,
\be
S^{\prime \prime}(\Psi^{(0)}) \xi = \pi^{(0)}\Lap \Psi^{(0)} + 
Q^T \Cav^\dagger \lambda
\ee
The lagrange multiplier $\lambda$  ensures the constraint
 (\ref{xiConstr0}). It
is a standard result known from the work of Kupiainen and Gawedzki
\cite{GK80} that the 
solution of such a linear equation can be written in the form
\ba
\xi &=& \Gamma \pi^{(0)} \Lap \Psi^{(0)}, \\
\Gamma &=& \lim_{\kappa \mapsto \infty} ( S^{\prime \prime } +
\kappa Q^T\Cav^\dagger \Cav Q )^{-1}  
\ea 
$\Gamma$ agrees with the full high frequency propagator in the background field
 $\Psi^{(0)}$.

 Next we recall the fact, recorded in section  {\ref{sec:1loop},
 eq.(\ref{GammaAlt1})
 that the full high frequency propagator
in a background field $\Psi $  agrees with 
$\pi \Gamma_{Q[\Psi]}\pi $ to zeroth order
 in $\nabla\Psi $. To the desired accuracy we can therefore replace
 $\Gamma $ by $\pi^{(0)}\Gamma_Q$. 

We record the final result for the background field
\ba
\Psi &=& (\Psi^{(0)} + \xi)/\mathrm{(modulus)} , \\
\xi &=& \pi^{(0)} \Gamma_{Q[\Psi^{(0 )}]} \pi^{(0)} \Lap \Psi^{(0)}, \\
\pi^{(0)}(z) &=& 1 - \Psi^{(0)}(z)\Psi^{(0)T}(z) \label{psi0}
\ea
with $\Psi^{(0)} $ from subsection \ref{subsec:psi0} below.
The division by the modulus is to ensure exact validity 
of $|\Psi|^2 =1$; note that $\mathrm{(modulus)} = 1 + O(\xi^2)$. 

Note that small discontinuities of normal derivatives of $\Psi^{(0)}$ at 
block boundaries give small contributions. On kinematical grounds, the 
derivatives are proportional $s^{-1}$ (cp. below) times a small factor 
if $\Phi $ is smooth. On the other hand, the length of the 
boundary is proportional to $s$ for blocks with $s^2$ lattice points.  

Therefore $\xi $ will be small if $\Phi $ is reasonably smooth.
\subsection{Smooth interpolation of blockspin fields}
\label{subsec:psi0}
Here we seek a field $\Psi^{(0)} (z)$ on the continuum which has a given 
block spin
 \be \Phi (x) = \av{z\in x}\Psi (z) / (modulus) \equiv  \Cav \Psi (x)/ (modulus),
\nn \ee
 which is 
continuous and smooth except for (small) discontinuities of the normal  derivative
on block boundaries, and which is close to $\Phi (x)$ for $z\in x$ if 
$\Phi $ is smooth.  
The average $\av{} $ is over the {\em lattice points} inside the square. 

 The lattice field $\Psi^{(0)} $ is  obtained by restriction 
to points $z$ in the lattice. 

We assume that the block spin $\Phi $ is reasonably smooth so that 
\hfill \\ 
\mbox{$\Phi (x)\cdot \Phi (y) >0 $} when $x,y$ are nearest or next nearest
 neighbours. This restriction removes some sign arbitrariness which could 
otherwise  lead to discontinuities. 

The continuum is divided into squares $x$ of sidelength $\tilde a$;
the lattice points inside form a block. 
 
We proceed in several steps. 

1. We determine the field at the corners $z_c$ of the squares,
\be
\Psi^{(0)} (z_c) = (\sum_x \Phi (x))/(modulus) \ . \label{corner}
\ee
where the sum goes over the four squares $x$ with corner $z_c$.

2. We consider the interpolations of the values of the function at 
the corners to functions on the sides between two adjacent corners. 
In this way, $\Psi^{(0)} $ is defined on the whole boundary of every square $x$,
and is close to $\Phi (x)$ when $\Phi $ is smooth.

Consider the side  with endpoints $z_0$ and $z_1$ which separates squares $x$ and $y$. Let
the 4 squares with joint corner $z_0$ be $x,y,x_0,y_0$ and the squares with joint corner
$z_1$ be $x,y,x_1,y_1$. 
If the side  is parametrized by $t=0...1$, with $z_0=z(0)$, $z_1 = z(1)$,
the interpolation is  as follows. 
\ba \Psi^{(0)}(t) &=& \left(   
\Phi (x) + \Phi (y) \right. \label{side} \\ 
&&\left. \quad  + (1-t)[\Phi (x_0) + \Phi (y_0)] + t[\Phi (x_1) + \Phi (y_1)]\right)
/(modulus) \nn
\ea

3.  We consider one square $x$ at a time and construct a 
preliminary interpolation  $\tilde \Psi^{(0)} $ which interpolates
 $\Psi^{(0)} $ from 
the boundary to the inside, such that it is smooth inside and takes 
the prescribed values on the boundary. The resulting function on the whole 
continuum is smooth except for discontinuities of the normal derivatives
across the boundaries of the squares. The interpolation is as follows.
Introduce the notation
$$  \Psi^\perp_x(z) = \Psi^{(0)} (z) - \Phi (x) (\Psi^{(0)} (z)\cdot \Phi (x)) $$
etc. 
Given $\Psi^\perp_x (z)$, the field $ \Psi^{(0)} (z)$ for $z\in x$ can be recovered
by eq.(\ref{parallel}) below. 
\footnote{In general, 
$ \Psi^\perp_x(z) \not=  \Psi^\perp_y(z) $ for $z$ on the side separating 
squares $x$ and $y$, because of the jump of $\Phi$. 
This is why a linear interpolation of $\Psi^\perp $ could not be used on the
 sides. 
Note however that there are no lattice points on the sides; 
every lattice point belongs to a unique square. }

Let the points $z$ in the closed square $x$ be parametrized by 
$(t_1, t_2)$, $0 \leq t_i \leq 1$. The four sides of the square have 
$t_1 = 0$ or $t_1=1$ or $t_2=0$ or $t_2=1$ respectively.
Regard $\Psi^\perp_x $ etc. as a function of $(t_1, t_2)$.
We consider first the linear interpolation $\tilde \Psi^\perp$ of the boundary values of 
$\Psi^\perp $ to the inside of the square,
\ba 
\tilde \Psi^\perp(t_1,t_2) 
&=&  (1-t_1) [\Psi^\perp(0, t_2) - \frac 12 (1-t_2) \Psi^\perp (0,0)
- \frac 12 t_2 \Psi^\perp (0,1)  ] \nn \\
&& + t_1 [\Psi^\perp (1,t_2) - \frac 12(1-t_2) \Psi^\perp (1,0)
- \frac 12 t_2 \Psi^\perp (1,1)] \nn \\
&& + (1-t_2) [\Psi^\perp (t_1,0) - \frac 12 (1-t_1) \Psi^\perp (0,0) 
- \frac 12 t_1 \Psi^\perp (1,0) ] \nn \\
&& + t_2 [\Psi^\perp (t_1,1) - \frac 12 (1-t_1) \Psi^\perp (0,1)
- \frac 12 t_1 \Psi^\perp (1,1) ]. \label{interior}
\ea

4. Adjust the value of the block spin while retaining the values 
of $\Psi^{(0)}$ at the boundaries and maintaining the smoothness. 
Again this is done separately for the squares $x$, using local 
coordinates $(t_1,t_2)$  as above. 
\ba  \Psi^\perp_x (t_1,t_2) &=& \tilde \Psi^\perp_x (t_1,t_2) -
 \alpha k(t_1, t_2),  \label{adjust} \\
k(t_1,t_2)&=& \sin (\pi t_1)\sin ( \pi t_2)  \nn \\
\alpha &=& s^2 \sin^2\left( \frac {\pi}{2s}\right) \Cav \tilde \Psi^\perp_x (x), 
\nn
\ea
if there are $s^2$ lattice points per square. 
 $\Cav $ takes the average over the lattice points
 inside the square $x$ similarly as before. 
The real function $k$ vanishes at the boundaries of the square.
Its  block average  is  $[s \sin (\pi / 2s)]^{-2}$. Therefore
$ \Cav \Psi^\perp_x (x) = 0$ as desired. 

 The field $\Psi^{(0)} $  is determined from $\Psi^\perp_x $,   
\be \Psi^{(0)} (z) = \Psi^\perp_x (z) + 
\left(1-|\Psi^\perp_x(z)|^2\right)^{\frac 12 } \Phi (x).
\label {parallel}\ee 
The positive square root is understood.
 The result satisfies all the requirements. Since the z-coordinates are 
$z_\mu = a s t_\mu + z_\mu^0$  where $z^0_\mu $ are the coordinates of the 
lower left corner of square $x$, the discontinuities in $\nabla \Psi^{(0)} $ across 
boundaries are of order $s^{-1}$ if the blocks are large. 

Let us note the locality properties of the construction. For $z\in x$,
$\Psi^{(0)}(z)$ depends only on the value of the blockspin $\Phi$
 at $x$ and at the 
8 nearest and next nearest neighbours of $x$. 

$\Psi^{(0)}$ is an explicitly 
given function of these 9 values by virtue of the formulas above. It is 
a nonpolynomial function of $\Phi(\cdot)$ because of the factor $1/(modulus) $
in eq.(\ref{side}) and  the 
factor with the square root in eq.(\ref{parallel}).
But if $\Phi $ is sufficiently smooth, these factors
could be expanded to obtain a polynomial approximation. 
\subsection{Vanishing of the correction term to Polyakov's result in order
$ \ln(a^\prime/a)$}
\label{sec:polyakovCorr}
Our result 
for the effective action appeared not to agree exactly with Polyakovs 
result to order $\ln (a^\prime/a)$. There is to this order a correction term
 \be
\beta v_M(0)\int_z [cos \theta (z) ]^{-1} \Phi^\perp \Lap \Psi (z) 
\ee 
Here we wish to show that this term is actually $0$ as a consequence of 
the extremality condition on the background field. 

{\bf Remark:}
{\em There is a very small remainder in the exact result because
$v_M(0)$ gets replaced by a matrix $\Gamma_Q (z,z) $ which is not 
diagonal in order $\hat{Q} -1$. However, because $\Phi^\perp $ is also  small,
this term is negligible in first order in $\hat{Q}-1$. }

To derive the result,
 we need the equation for the background field $\Psi $ in a form
 which was not used before.  

 General fields $\phi $ which  satisfy the block spin
 constraint  can be parametrized in terms of a field  
$\zeta (z) \perp \Phi (x)$ which satisfy the block spin constraint 
\be \Cav \zeta = 0 \label{czeta}\ee
according to eq.(\ref{phiZeta}). For notational simplicity introduce
\be
\tilde \Phi (z) = \Phi (x) \ \mbox{for } z\in x. \label{PhiTilde}
\ee
To first order in the deviation of
 $\phi $ from $\Psi $,
\be
\phi = \Psi  + \zeta- \frac {1}{\Psi \cdot \tilde \Phi } (\Psi \cdot \zeta) \tilde \Phi 
\ee
We may abandon the constraint $\zeta (z)\perp \tilde \Phi (z) $ 
because a component $\zeta^0(z) \tilde \Phi (z)$ of $\zeta$ 
in the direction of $\tilde \Phi (z)$ contributes nothing to $\phi $. 
The extremality condition reads therefore 
\ba
\zeta^T \hat{S}^\prime (\Psi) = 0, \label{extrS} \\ 
\hat{S}^\prime (\Psi )= \frac {\delta }{\delta \zeta }S(\phi )|_{\phi=\Psi}. \nn
\ea
for $\zeta $ which satisfy constraint (\ref{czeta}).This is equivalent to 
\be
\hat{S}^\prime (\Psi ) = \Cav^\dagger \lambda 
\ee
with a Lagrange multipliers $\lambda $. $\lambda $ is a field on the 
coarse lattice.

Working out the derivative of $S$ we find the nonlinear equation
\be
\Lap \Psi -  \frac {\tilde \Phi \cdot \Lap \Psi }{\Psi \cdot \tilde \Phi } \Psi =
\Cav^\dagger \lambda .  \label{eqPsi}
\ee
 Since
 $\tilde \Phi$  is constant on blocks, 
it follows from eq.(\ref{eqPsi}) that 
\be
\tilde \Phi \Cav^\dagger \lambda = \Phi(x) \lambda(x)=0 \label{lambdaPerp}
\ee
By definition $\tilde \Phi^\perp \cdot \Psi = 0. $
Therefore 
\ba
- \tilde \Phi^\perp \Lap \Psi&=& - \tilde \Phi^\perp \left(
-\Lap \Psi  + \Psi \frac {\tilde \Phi \Lap \Psi}{\tilde \Phi\cdot \Psi }\right) \\
&=& \tilde \Phi^\perp \Cav^\dagger \lambda
\ea
In the second equation, eq.(\ref{eqPsi}) was used. 

Inserting the definition of $\Phi^\perp$, we compute
\ba
\int_z [cos \theta (z) ]^{-1} \Phi^\perp \Lap \Psi (z)  &=& 
\int_z \frac {1}{\cos \theta (z)} \left[
\tilde \Phi -  \Psi (\tilde \Phi \cdot \Psi ) \right](z) (\Cav^\dagger \lambda )(z)
\nn \\
&=& \int_x \Cav \left( \frac{1}{\cos \theta } \tilde \Phi - \Psi \right)(x) 
\lambda (x). \label{int_x}
\ea
We will show that the integrand in expression (\ref{int_x}) is zero. 
This shows that the correction term is zero.

$\Cav \Psi (x) = \rho (x) \Phi (x) $ by the block spin definition,
with some real $\rho (x)$. Moreover, because $\tilde \Phi $ is 
constant on blocks, it follows that   
\be
\Cav \left( \frac {1}{\cos \theta}\tilde \Phi  \right)(x) =
\Phi (x) \av{z\in x} \left( \frac {1}{\cos \theta}\right)(z).
\ee
This is again a multiple of the vector $\Phi (x)$. But according to 
eq.(\ref{lambdaPerp}), \mbox{$\Phi(x)\cdot \lambda(x) = 0$}. Therefore the 
integrand in expression (\ref{int_x}) vanishes, and the result is proven.

\section{Gaussian block spin} 
\label{sec:gaussian}

 Define the linear averaging
 operator $\Cav^\perp[\Phi ] $ which depends parametrically on $\Phi $ by 
\be
\Cav^\perp [\Phi ] \phi (x) = \Cav \phi (x) - \Phi(x)
 (\Phi (x) \cdot \Cav \phi (x)) = \Cav \phi^\perp (x).
\ee
 The $\tilde \kappa $-dependent effective action  is 
defined as follows.
\be
  \label{weactKappa}
e^{-S_\mathrm{eff}[\Phi]}=
        \int\,\mathcal{D}\phi\prod_x \left(
J_0(\Cav \phi (x)) 
e^{-\frac {\tilde \kappa}{2} ||\Cav^\perp[\Phi] \phi (x)||^2} \right)
e^{-S[\phi]} \ ,
\ee
where $J_0$ is a $\tilde \kappa $-dependent jacobian which ensures that  
\be
\int \mathcal{D} \Phi \prod_x 
\left( J_0(\Cav \phi (x))
e^{-\frac {\tilde \kappa} {2} ||\Cav^\perp[\Phi] \phi (x) ||^2}
\right)  = 1 \label{J0Kappa}
\ee
for all $\phi $.  Explicitly (see Appendix A)
\ba
J_0 (\Xi )^{-1} &=& \int \frac {d^{\mathcal{N}}\pi}{\sqrt{1 - |\pi |^2}} 
e^{-\frac {\kappa}{2} |\Xi|^2 |\pi |^2}   \nn   \\  
&=& const \cdot \left( |\Xi|^2  - \frac {1}{\tilde \kappa} + ...\right)^{-\mathcal{N}/2}
\label{J0Result}
\ea
The last formula is valid for large $\tilde \kappa$. 

If it is the aim to improve the locality properties of the 
classical perfect action, one should choose $\tilde \kappa $ of order
$\beta$,
\be
\tilde \kappa = \beta \kappa \tilde a^2.
\ee
One introduces a background field $\Psi (z)$ which extremizes the exponent, viz
\be
S(\phi ) + \frac {\tilde \kappa}{2} \sum_x |\Cav \phi^\perp (x)|^2  = Extr. 
\label{extrKappa}
\ee  
at $\phi = \Psi$. Then one parametrizes the field $\phi $ in terms of a
fluctuation field $\zeta $ as before, viz. $\phi^\perp = \Psi^\perp + \zeta$.
It follows that 
\ba
a^2 \sum_x |\Cav \phi^\perp (x)|^2 &=& \int_x 
|\Cav \Psi^\perp (x)|^2 + \int_z\left[ 2 \zeta \cdot \Cav^\dagger \Cav \Psi^\perp (z)
+ \zeta \cdot \Cav^\dagger \Cav \zeta (z)\right] \nn
\ea 
We make the transition to $\xi$-variables  $\xi = Q^{-1}\zeta$ and expand in 
powers of $\xi $. By eq.(\ref{extrKappa}), 
the term linear in $\xi$ must vanish. Putting $\tilde \kappa = \beta \kappa \tilde a^2 $
we obtain the saddle point condition 
\be
S^\prime (\Psi ) =  -\beta\kappa Q^T\Cav^\dagger  \Cav \Psi^\perp ,
\label{saddleKappa}
\ee
and the classical perfect action, which is the value of expression 
(\ref{extrKappa}) at the extremum $\xi = 0$, comes out as
\be
S_{cl}[\Phi ] = S(\Psi) + \frac {\beta \kappa}{2}\int_x| \Cav \Psi^\perp (x)|^2.
\label{SclKappa}
\ee 
Our definition of the classical perfect action does not include the jacobian. 
The classical perfect action is of order $\beta$, while the logarithm
of the jacobian $J_0$ is of order $\beta^0$. 

$\Cav \Psi^\perp (x)$ comes out to be of order $1/\kappa $. Therefore the
second term in eq.(\ref{SclKappa}) vanishes in the limit 
$\kappa \mapsto \infty $ and we recover the previous result.

From here on the calculation proceeds exactly as before, 
and the result is the same as for $\kappa = \infty$, except for the 
following changes
\begin{enumerate}
\item
The jacobian $J_0$ now has the form (\ref{J0Result}) which contains a mild 
$\kappa$-depen\-den\-ce. 
\item The background field is determined by the new saddle point condition
(\ref{extrKappa}). 
\item The classical perfect action is given by eq. (\ref{SclKappa}). 
Apart from the change of the background field, there is an extra term in it. 
\item The high frequency propagators $\Gamma $ and 
interpolation operators  $\Aop $ with finite $\kappa $ have to be used 
throughout. 
\end{enumerate}
The background field will be examined below. The result is that our
previous analytical approximation for $\Psi $ remains valid for large 
enough $\kappa$, except that the high frequency propagator with finite 
$\kappa $ has to be substituted.
 A ``small $\kappa$ approximation'' will also be mentioned.
Both approximations become exact when the blockspin field tends to a constant.

\subsection{The background field for finite $\kappa$}   
One should solve eq.(\ref{saddleKappa}).
Suppose that an approximate solution $\Psi^{(0)} $ is at hand. Then an
improved solution $\Psi = \Psi^{(0}) + \xi + O(\xi^2)$ is determined as 
before in section \ref{sec:background}. 
Expanding to first order in $\xi $, eq.(\ref{saddleKappa})
takes the form
$$ 
\left[S^{\prime\prime}[\Psi^{(0)}] + \beta \kappa Q^T \Cav^\dagger \Cav Q  \right] \xi
= - S^\prime [\Psi^{(0)}] - \beta \kappa Q^T \Cav^\dagger \Cav \Psi^{(0)\perp }
$$
with approximate solution 
\be
\Psi = \left( \Psi^{(0)} + \pi^{(0)}\Gamma_{Q[\Psi^{(0)}]} \pi^{(0)}
[\Delta \Psi^{(0)}  - \beta\kappa  
Q^T Q \Cav \Psi^{(0)\perp } ] \right) / (modulus) + O(\xi^2), \label{psiKappa} 
\ee
where $\Gamma_Q$ is the high frequency propagator (\ref{GammaQ}) with finite
$ \kappa $.  

A zero approximation $\Psi^{(0)}$ can be constructed in the same way as in 
section \ref{sec:background},
possibly with a different choice of the vector $\alpha $.

We consider two choices
\begin{description}
\item [large $\kappa$ approximation.] We choose $\alpha $ as before, so that
$\Cav \Psi^{(0)\perp}   = 0.$ Then the $\beta\kappa$-term in 
eq.(\ref{psiKappa})  vanishes and we obtain the same fromula for $\Psi $ as 
before, except for the use of the finite-$\kappa$-propagator.
\item[small $\kappa$ approximation.] Choose $\alpha =0 $ and use the 
full eq.(\ref{psiKappa}).
\end{description}
Let us now discuss why the effective action with suitable finite $\kappa$ is
expected to have better locality properties than at $\kappa=\infty$. 
This comes out of the better falloff properties of the 
high frequency propagators $\Gamma$ and the interpolation operators $\Aop $.
 These locality properties  are
inherited by the perfect classical action. And the corrections to the 
perfect 
classical action also benefit from the improved falloff properties of 
$\Gamma $ and $\Aop$. 

The falloff properties  of $\Gamma_Q$, which appears in the analytic formula 
for the 
background field, are inherited from those of $\Gamma_{KG}$. Since
$\Aop_{KG} = \kappa \Gamma_{KG}\Cav^\dagger $, this follows from the
 perturbation expansion  of $\Gamma_Q $ in powers of $\hat Q -1$. 

In conclusion, if one wishes to 
achieve good locality properties in the effective action,
 $\kappa $ should be so chosen that 
$\Gamma_{KG}$ has good locality properties. As we said 
in the introduction, this has a prize. Systematic tests of the 
accuracy of various approximations are easier with $\kappa = \infty$.


\section*{Acknowledgement}
Work supported in part by Deutsche Forschungsgemeinschaft. 
G.\,P. was partially supported through projects \emph{FONDECYT}, Nr.\,1980608,
 and  \emph{DICYT}, Nr.\,049631PA.
He would like to thank the II. Institut for Theoretical Physics of the University
 of Hamburg for the kind hospitality.

\section*{Appendix A: The Jacobian}

The effective action is defined by eq. (\ref{weact})
\ba
e^{-S_\mathrm{eff}[\Phi]}&=&
        \int\,\mathcal{D}\phi\prod_x\delta(C\phi(x),\Phi(x))e^{-S[\phi]};\\
\mathcal{D}\phi &=&\prod_z d\phi(z) .
\ea
The argument of the $\delta$-function is nonlinear. 
But the blockspin definition $\Phi (x)=\frac{\Cav \phi (x)}{|\Cav \phi (x)|}$ 
is equivalent to $\Cav \phi(x) - \Phi(x) \left(\Cav \phi(x) \cdot \Phi(x) \right)=0$.
Using the linear block average (\ref{Cav}) and the parametrization (\ref{phiZeta}) 
we finally end up with the linear condition $\Cav \phi^\perp (x)=\Cav \zeta (x) =0$.
Now we want to compute the jacobian $J(\Psi,\zeta)$
associated to the parametrization which leads
to a linear condition.
The blockspin definition is implemented by $\delta$-functions which are centered on blocks
\be
\prod_x\delta(C\phi(x),\Phi(x)).
\ee 
Therefore we choose a local basis with
\be
\Phi (x)=\left( 1,0,\dots,0\right).
\ee
We denote by $F^i$ the components of the new blockspin condition
\be
F^i=\Cav\phi^i-\Phi^i \left(\Cav \phi \cdot \Phi \right)=0,\qquad i=1,\dots \mathcal{N}.
\ee
In the following, we neglect to write arguments $x$. 
The jacobian is given by 
\be
J_0(\Cav \phi)=|\mbox{det}\frac{\delta\,F^i}{\delta\,\Phi^j}|(\phi).
\ee
We compute
\be
\frac{\delta\,F^i}{\delta\,\Phi^j}(\phi)=-\delta^{ij}\left( \Phi\cdot\Cav\phi \right)
\ee
and find
\ba
J_0(\Cav \phi)&=&e^{\mathcal{N}\ln \Phi\cdot\Cav\phi } \nn \\ 
&=&| \Phi\cdot\Cav\phi|^\mathcal{N}
\ea

Consider now the Jacobian  $J_0 $ for finite $\kappa $ as defined by 
eq.(\ref{J0Kappa}). 

$\Cav^\perp [\Phi]\phi (x)$ depends on 
$\Phi $ only through $\Phi (x)$. Let us write $\Phi $ in place of the variable
$\Phi (x)$ in the following. We must compute
$$
J_0(\Cav\phi (x))^{-1} = \int
d\Phi e^{-\frac \kappa 2 ||\Cav^\perp[\Phi] \phi (x) ||^2}. 
$$ 
Let $\Xi = \Cav \phi (x)$. Then 
$$
||\Cav^\perp[\Phi] \phi (x) ||^2 = |\Xi|^2 - |\Phi \cdot \Xi |^2= |\Xi|^2 |\pi^2|  
$$
if we choose a basis so that $\Xi $ points in 0-direction, and
write $$\Phi = (\sqrt { 1 - |\pi|^2}, \pi)\ .$$
 Using the standard representation 
of the uniform measure $d\Phi $ on the sphere in terms of coordinates $\pi$, 
we get expression (\ref{J0Result}). In the limit $\kappa \mapsto \infty$, 
 the result agrees with the formula given above.    
\subsection*{Appendix A.1 $\det Q$}
We also need the jacobian of the transformation from $\zeta$-variables 
to $\xi$-variables. The integration variables $\xi^i$ are the coefficients
of $\xi$ in an orthonormal basis $(e_1,\dots, e_{\mathcal{N}})$ for 
the tangent space $\TSz $. Such a basis comes from an orthonormal basis
for $ \RNP $ with $e_0= \Psi(z)$. Similarly, the integration variables 
$\zeta^i $ are  the coefficients
of $\zeta$ in an orthonormal basis $(f_1,\dots, f_{\mathcal{N}})$ for 
the tangent space $\TSx , \ x \ni z$.
 Such a basis comes from an orthonormal basis
for $ \RNP $ with $f_0= \Phi(x)$. Since $Q\xi = \zeta$, 
\be
\prod_1^\mathcal{N} d\zeta^k = |\det (Q^{i}_j)| \prod_1^\mathcal{N} d\xi^k   
\qquad \mbox{ if } \ Qe_j = \sum_{i=1}^\mathcal{N}f_i Q^{i}_j. 
\ee
Unwritten arguments are $z$. 
 
The modulus of the determinant is independent of the choice of orthonormal 
bases since $\det O=\pm 1$ for orthogonal transformations $O$. 
 Therefore we may choose convenient bases as follows. 
\be e_1 = \Phi^\perp (z)/|\Phi^\perp (z)| = \frac 1 {\sin \theta (z)}
\left[ \Phi (x) - \Psi (z) (\Psi (z)\cdot \Phi (x)\right],
\ee
and $e_2,...,e_\mathcal{N}$ an arbitrary completion to an 
orthonormal basis. Similarly we choose
\ba
f_1 &=& -\Psi^\perp(z)/|\Psi^\perp(z)| =  
\frac {-1} {\sin \theta (z)}
\left[ \Psi (z) - \Phi (x) (\Psi (z)\cdot \Phi (x)\right], \nn \\
f_k &=& e_k \ \mbox{ for } \ k=2,\dots,\mathcal{N}.
\ea
Basis vectors $f_0,f_1$ are linear combinations of $e_0, e_1$. Therefore 
$e_k, k=2,\dots,\mathcal{N}$ are orthogonal to them and the basis vectors 
$f_i$ are indeed orthonormal. Using eq.(\ref{Qdef}) we compute 
\ba
 Qe_1&=& \cos \theta(z) f_1 , \nn \\ 
 Qe_k &=& f_k \ \mbox{for} \ k=2,\dots,\mathcal{N}. 
\ea
Thus, the matrix $(Q^{i}_j)$ is diagonal with a single eigenvalue 
$\cos \theta (z)$ which is distinct from 1. Therefore
\be
|\det (Q^{i}_j) | = \cos \theta (z).
\ee

\section*{Appendix  B: The kinetic term}
Let us write
\be
\label{kineticTerm}
|\nabla_\mu \xi |^2 + | \nabla_\mu \xi^0 |^2 = |\nabla_\mu \varphi|^2 + 
\Delta L_{\mathrm{kin}}
\ee
Our task is to evaluate $\Delta L_{\mathrm{kin}}$. It turns out to be of order
$|\nabla_\mu \Psi |^2$. Therefore it will later be treated as a perturbation
which needs to be taken into account to  first order  only. 

{\bf Conventions:}
{\em Arguments not written are $z$. 
To save brackets, we agree that derivatives $\nabla_\mu $ act only on the
first factor behind them.  
 }

We use the exact lattice Leibniz rule 
(\ref{Leibniz}) throughout. It turns out that this is
essential. 

By definition, $\xi = \pi \varphi $ and $\xi^0 = \Psi^T \varphi $, 
while $ \pi = 1- \Psi \Psi^T$.

The use of the lattice Leibniz rule is slightly subtle, because there are always two ways to use it which differ by the assignment of which factor is $f$ and which is $g$. Choices have to match, so that factors in $\Psi^T\varphi$ 
have the same argument, $z+\hat \mu$, and factors in $\Psi^T \Psi $ also have
 the same argument, $z$. Apart from this the calculation is straightforward and gives

\ba
\Delta L_{\mathrm{kin}} &=&
\varphi^T(z+\hat \mu)\left[ 2 \nabla_\mu \Psi \nabla_\mu \Psi^T    
+ \Psi (z+\hat \mu )\nabla_\mu \Psi^T \nabla_\mu \Psi \Psi^T (z+\hat \mu ) \right]\varphi (z+\hat \mu) \nn
\\ & & + ( \nabla_\mu \varphi^T j_\mu \varphi (z+\hat \mu )+ \mbox{transpose}) \ . 
\ea
Because of the smoothness of $\Psi $, we are only interested in terms 
up to order $|\nabla \Psi|^2$. This has been used to 
approximate $\Psi^T(z+\hat \mu)\Psi (z) = 1 + \mbox{negligible}$
when multiplied with two factors $\nabla \Psi $. 
\section*{Appendix C: The second order term}
\label{sec:appendixC} 
We wish to evaluate the quantity
\be
I = \langle \nabla_\mu \varphi^T(z)j_\mu(z)\varphi(z) \
\nabla_\mu \varphi^T(w)j_\mu(w)\varphi(w) \rangle^T
\ee
There are two possible contractions and we obtain
\ba
I(z,w) &=& I_1+I_2 \\
I_1 &=& 
\tr \left( \nabla_\mu \Gamma_Q(z,w)\leftNabla_\nu j_\nu(w)\Gamma_Q(w,z)
j_\mu^T(z) \right)  \\ 
I_2&=& \tr \left( j_\mu(z) \Gamma_Q(z,w)\leftNabla_\nu j_\nu (w)
\Gamma_Q(w,z) \leftNabla_\mu\right) 
\ea
We wish to extract the singular part which is proportional to $\ln \tilde{a}/a$
since this part will contribute to the Polyakov result. 
Using $j_\mu(z) = j_{-\mu}^T(z+\hat{\mu})$
$(\approx - j_\mu^T(z))$  we see by partial integration that  
the singular part of $I_1$ and $I_2 $ are equal. 

We need the singular part of the 1-loop Feynman graph
\be
{\cal G}_{\mu \nu}^{\beta \gamma , \delta \epsilon}(z,w) =
\nabla_\mu \Gamma_Q(z,w)^{\beta \gamma}\leftNabla_\nu
 \Gamma_Q(w,z)^{\delta \epsilon} 
\ee
The singular part does not depend on $Q$ nor on details of the cutoff.
As in section \ref{PolyakovResult} we may therefore replace the 
propagators by Yukawa potentials $v_M (z-w){\bf 1}$ with mass $M$ 
of order $\tilde{a}^{-1}$. 

  By power counting and rotational invariance the singular part  must
 be of the form
$$G\delta_{\mu \nu }\delta^{\beta \gamma}\delta^{\delta \epsilon} \delta(z-w) $$
with an unknown coefficient $G$.
The coefficient can be computed by considering 
$\delta^{\nu \mu }{\cal G}_{\mu \nu}$.
 Since $-\Lap \Gamma = 1 + \mbox{nonsingular}$ 
it follows that $G=\frac 12$.

Inserting this yields
\ba
\int_w I(z,w) &=&  v_M(0)\tr j_\mu(z)^T j_\mu (z)   + \mbox{nonsingular}
\label{I1}
\ea
Since the singular parts of $v_M(0)$ and $\Gamma_{KG}(z+\hat \mu , z+\hat \mu)$
are equal, eq.(\ref{I1}) shows that the renormalized Feynman diagram 
$\Seff^{(2)}$ is indeed finite in the limit $a\mapsto 0$. 
\section*{Appendix D: Fourier transform \\ of Kupiainen Gawedzki kernels}
Given a lattice $\Lambda $ of lattice spacing $a$ with points $z,w,...$
and a block lattice $\tilde \Lambda $ of lattice spacing $\tilde a = sa$,
($s$ a positive integer) with sites $x,y,...$, 
we characterize the points by real coordinates 
$z^\mu$ resp. $x^\mu $ etc. The conjugate variables $k_\mu$ and $p_\mu $ take 
their values in the duals $\Lambda^\ast$ and $\tilde \Lambda^\ast$,
\ba &&-\frac \pi a < k_\mu \leq \frac \pi a ,     \\
&& -\frac \pi {\tilde a} < p_\mu \leq \frac \pi {\tilde a} . \label{boundFT}
\ea
If the lattices are infinitely extended, $p_\mu $ and $k_\mu $ are real
 variables. If the lattice $\Lambda$ has extension $La= \tilde L \tilde a$
instead,
\be p_\mu, k_\mu \in \frac {2\pi}{La} {\bf Z} \ee  
We use the notation $\int_k (...)  = \int d^2k $ if $\Lambda$
 is infinitely extended , and 
\be \int_k (\dots ) = \left(\frac {2\pi }{La}\right)^2 \sum_k (\dots )   \ee
otherwise. 
The same formulas is used for $\int_p$, only the boundaries of the 
integration are different according to eq.(\ref{boundFT}).  Let
\be D=\{ l | l_\mu \in \frac {2\pi}{\tilde a} {\bf Z},
\  -\frac {\pi} {a} < l_\mu \leq \frac \pi a \} \ . 
\ee
Then every $k\in \Lambda^\ast$ admits a unique decomposition 
\be k = p + l\ , \  p\in \tilde\Lambda^\ast,\ l\in D . \label{dec} \ee 

The Fourier transform of the massless lattice propagator $v(z-w)$ is
\be \tilde v (k) = \left(\sum_{\mu=1,2} \frac 2 {a^2}[1-\cos k_\mu a]\right)^{-1}
\ee
Because of invariance under translations by lattice vectors of the block
lattice, the averaging kernel $\Cav$, interpolation kernel 
$\Aop \equiv \Aop  _{KG}$, block propagator $u=u_{KG}$ and high frequency 
propagator $\Gamma \equiv \Gamma_{KG}$ admit Fourier expansions
\ba
\Cav (x,z) &=& (2\pi)^{-2}\int_{k\in \Lambda^\ast } 
\tilde \Cav(k) e^{ik_\mu(x^\mu - z^\mu)}\ , \\
\Aop (z,x) &=& (2 \pi)^{-2}  \int_{k\in \Lambda^\ast } 
\tilde \Aop(k) e^{-ik_\mu(x^\mu - z^\mu)}\ , \\
u(x-y) &=& (2\pi)^{-2} \int_{p\in\tilde \Lambda^\ast} \tilde u(p) e^{ip(x-y)} 
 , \\
\Gamma(z,w) &=& (2\pi )^{-2} 
\sum_{l,l^\prime \in D}\int_{p\in \tilde \Lambda^\ast}  
\tilde \Gamma_{ll^\prime}(p) e^{i(l^\prime z - lw)}e^{ip(z-w)}
\ea
The averaging kernel $\Cav(x,z) = \tilde a^{-2}$ for $z \in x$ and 
$0$ otherwise. Assuming $z\in x$ iff $- a/2 < z^\mu - x^\mu \leq a/2 $
one obtains
\be
\tilde \Cav(k) = \prod_{\mu=1,2} \left( \frac 2 {\tilde a k_\mu} \sin ( k_\mu a/2)
\right) 
\ee 
{\bf Notational convention: } {\em When variables $k, p$ appear
 together in one formula, they are related by the unique decomposition 
(\ref{dec}).}

From eqs.(\ref{KGkernels}) one computes 
\ba
\tilde u(p) &=& \sum_{l\in D} \tilde v (p+l) |\tilde \Cav (p+l)|^2 \ , \\
\tilde \Aop(k) &=&  \tilde v (k) \tilde \Cav(-k) \tilde u(p)^{-1} \ , \\
\tilde \Gamma_{ll^\prime}(p) &=& \delta_{ll^\prime} \tilde v(p+l) - 
\tilde \Aop(p+l^\prime)\tilde \Cav(p+l) \tilde v (p+l) \ . 
\ea
The variables $l,l^\ast $ assume $s^2$ values each. 

In $d$ dimensions, the formulas remain valid, except that
$\mu = 1,...,d$, and factors
$(2\pi)^d $ have to be substituted for $(2\pi)^2$.
The variables $l,l^\prime$ now assume $s^d$ values.

\end{document}